%% file: main.tex
\documentclass[12pt]{article}

\usepackage{enumerate}

\usepackage{amsmath, amsthm, amssymb, bm, natbib,tikz}
\usepackage{framed,xcolor,geometry}
\usepackage{graphicx}
\usepackage{threeparttable,booktabs,lscape,pdflscape,multicol, url}

\usepackage{multirow,rotating, tabularx}

\usepackage{verbatim,epstopdf}
\usepackage{caption}
\usepackage{fancybox,url}

\def\bSig\mathbf{\Sigma}
\DeclareMathOperator{\df}{df}

\newcommand{\blind}{0}

\renewcommand{\baselinestretch}{1.6}

\input{macros}

\date{}
\begin{document}
\definecolor{shadecolor}{gray}{0.9}

\if0\blind
{
    \title{\Large A Unified Framework for  Nonlinear Mediation Analysis \\ of  Random Objects}
    \author{Wenxi Tan, Bing Li, and Lingzhou Xue \\ Department of Statistics, The Pennsylvania State University}
  \maketitle
} \fi

\if1\blind
{
  \title{\Large A Unified Framework for Nonlinear Mediation Analysis \\ of Random Objects}
    \author{}
  \maketitle
} \fi

\begin{abstract}
Mediation analysis for complex, non-Euclidean data, such as probability distributions, compositions, images, and networks, presents significant methodological challenges due to the inherent nonlinearity and geometric constraints of such spaces. Existing approaches are often restricted to Euclidean settings or specific data types. We propose Random Object Mediation Analysis (ROMA), a unified framework that simultaneously accommodates object-valued exposures, mediators, and outcomes, enabling the analysis of nonlinear causal pathways in general metric spaces. ROMA leverages an additive Reproducing Kernel Hilbert Space (RKHS) operator model to rigorously disentangle direct and indirect causal pathways, which is a significant advancement over existing single-predictor or purely predictive additive frameworks. Theoretically, we establish the nonparametric identification of causal effects and derive global asymptotic normality for our estimators. Crucially, this theoretical foundation enables the construction of simultaneous confidence bands and global test statistics without the need for computationally intensive resampling. We demonstrate the practical utility of ROMA through simulations and real-world applications involving compositional mediators and distributional outcomes, extending the scope of mediation analysis.
\end{abstract}


\vfill

\section{Introduction}
Causal mediation analysis provides a principled framework for disentangling the mechanisms by which exposures influence outcomes through mediators, widely used in epidemiology, economics, and biomedical research \citep{richiardi2013mediation, guo2023highmed, song2020bayesian}. The total effect of an exposure is typically decomposed into two components: the natural indirect effect (NIE), which quantifies the impact transmitted through the mediator, and the natural direct effect (NDE), representing the pathway not mediated. The primary objective of causal mediation analysis is to identify, estimate, and infer these effects, thereby providing insight into the underlying causal mechanisms that drive observed associations.

\begin{figure}[htp]
    \centering
    \begin{tabular}{@{}cccc@{}}
    \includegraphics[width=0.35\linewidth]{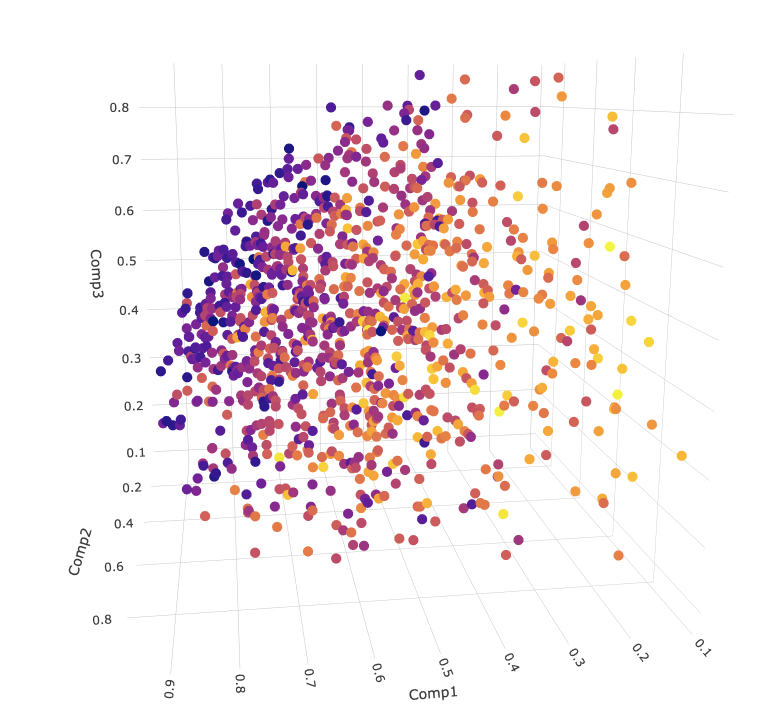} &
    \includegraphics[width=0.35\linewidth]{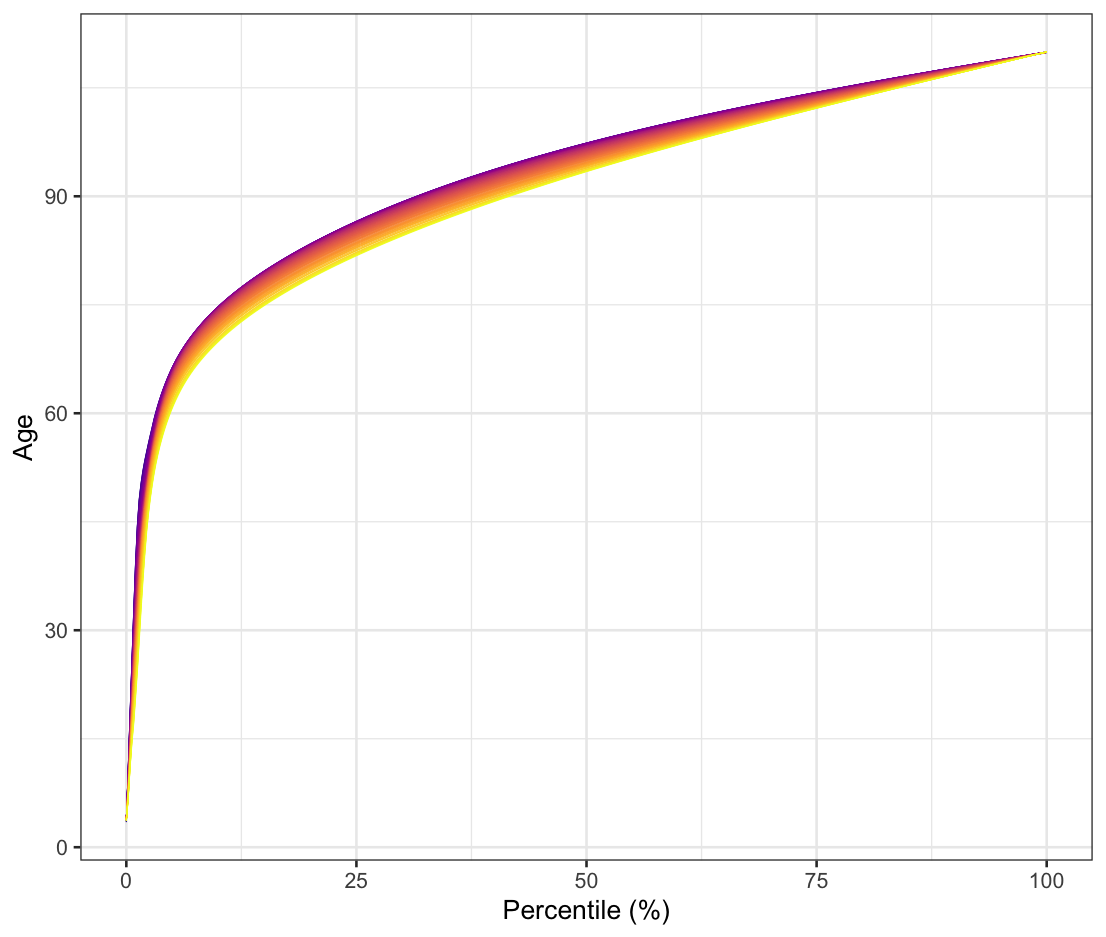}
\end{tabular}
  \caption{Examples of non-Euclidean outcomes varying with exposures. Left:  Compositional outcomes on the positive orthant of the unit sphere $\c S^2 \subset \R^3$. Right: Fitted quantile functions representing age-at-death distributions. The color gradient (blue to yellow) indicates increasing exposure intensity (temperature).}
    \label{fig: exp for outcomes}
\end{figure}

Traditional mediation analysis has been confined to linear spaces (specifically the Euclidean space $\R^p$) and is typically formulated using linear structural equation models (LSEMs). Recent methodological advances have been largely restricted to cases where the mediator is a specific type of random object, such as probability densities \citep{zhangetldistM}, microbiome compositions \citep{CompoMedianHongzhe}, or functional data \citep{Coffman03092023}. These methods often rely on geometry-specific transformations, such as isometric log-ratios or functional principal components, that do not generalize to general metric spaces. Importantly, these approaches are typically limited to scalar or Euclidean outcomes, failing to capture the full complexity of the causal chain when the outcome itself is a random object. Moreover, statistical inference in these approaches typically lacks a unified asymptotic framework, necessitating computationally intensive resampling procedures (e.g., bootstrap) to quantify uncertainty. The practical necessity of a unified framework addressing such complex structures is highlighted in Figure \ref{fig: exp for outcomes}, which illustrates how both compositional and distributional outcomes exhibit complex, non-Euclidean variation with exposure.

Our research is specifically driven by an environmental health application designed to disentangle the mechanisms linking extreme temperature, air pollution, and mortality. While both exposures are associated with adverse health outcomes \citep{zanobetti2008temperature, wu2020air}, and air pollution is known to modify the effect of heat waves \citep{analitis2014effects}, the role of pollutants as mediators remains less understood. Prior studies typically collapse age-at-death distributions into scalar mortality rates, a simplification that often fails to capture complex, age-specific mortality patterns. In contrast, we investigate this mediation mechanism by treating the full age-at-death distribution as the outcome, allowing for a more granular understanding of the mediation pathway.

Recent years have witnessed substantial advances in the statistical analysis of random objects. Foundational work by \cite{petersen_frechet_2017} established the global linear Fr\'echet regression for random objects with Euclidean predictors, which \cite{zhang2024dimension} augmented with sufficient dimension reduction techniques. Expanding the scope beyond linearity, \cite{bhattacharjeeNonlinearGlobalFrechet2023} introduced the global nonlinear Fr\'echet regression by introducing the notion of a weak conditional Fr\'echet mean via reproducing kernel Hilbert space (RKHS) embeddings, enabling models with both object-valued responses and predictors. Parallel developments have emerged in causal inference for random objects. \cite{linCausalInferenceDistribution2023} characterized the causal effects with optimal transport maps in the Wasserstein space of the distributional outcomes,  \cite{kurisu_geodesic_2024} examined outcomes residing in geodesic spaces, and \cite{bhattacharjee_doubly_2024} presented a unifying Fr\'echet embedding principle to facilitate rigorous mean estimation and causal effect mapping. However, existing methods do not simultaneously accommodate object-valued outcomes and mediators, necessitating new methodological and theoretical tools for mediation analysis beyond the Euclidean setting.

To address these challenges, we propose Random Object Mediation Analysis (ROMA), a unified framework that simultaneously accommodates object-valued exposures, mediators, and outcomes and enables nonlinear causal pathways for random objects residing in metric spaces that admit a valid Hilbertian embedding (e.g., metric spaces of negative type). This class encompasses most spaces encountered in practice, including Wasserstein spaces, Fisher-Rao metrics, and compositional data geometries. Our approach leverages Fr\'echet embeddings to efficiently compute Fr\'echet means through expectations in a Hilbert space via Bochner integrals, effectively generalizing the weak conditional mean framework of \cite{li_dimension_2022} to the non-Euclidean mediation context. Crucially, we rigorously establish the nonparametric identification of causal effects in this non-Euclidean context, which is a foundational result that underpins our construction of flexible Hilbertian regression operators for modeling complex linear and nonlinear dependencies. Figure~\ref{fig: diag} presents the conceptual path diagram of our method, which we formalize following the construction of the RKHS kernels.

Theoretically, we provide a comprehensive inferential framework by establishing the convergence rates and asymptotic normality of the proposed estimators. A key advancement of our approach is that it enables the construction of pointwise confidence regions and global test statistics without the need for resampling, thereby overcoming a significant computational bottleneck in random object analysis. We validate the finite-sample performance through simulation studies and demonstrate the practical power of our methods in two real-world applications involving compositional mediators and distributional outcomes, respectively.

\begin{table}[ht]
    \centering
    \renewcommand{\arraystretch}{0.7}
    \caption{Comparison of the proposed ROMA framework with existing mediation methods.}
    \begin{tabular}{ccccc}
        \toprule
        \textbf{Method} & \textbf{Mediator} & \textbf{Outcome} & \textbf{Asymptotics} & \textbf{Inference} \\
        \midrule
        CCMM \citep{CompoMedianHongzhe} & Composition & Euclid & $\times$ & Bootstrap \\
        FunM \citep{Coffman03092023} & Function & Euclid & $\times$ & Bootstrap \\
        DistM \citep{zhangetldistM} & Distribution & Euclid & $\times$ & Bootstrap \\
        ROMA (Ours) & Object-valued & Object-valued & \checkmark & Analytical \\
        \bottomrule
    \end{tabular}
    \label{tab: varis}
\end{table}
Specifically, we summarize the contributions of our work as follows:
\begin{itemize}
    \item \textbf{A unified framework for nonlinear mediation with object-valued data.} Existing non-Euclidean mediation approaches typically focus on specific types of mediators (e.g., functional data \citep{Coffman03092023}, compositional data \citep{CompoMedianHongzhe}, probability distributions \citep{zhangetldistM} as listed in Table \ref{tab: varis}) and are often limited to linear relationships involving Euclidean exposures and outcomes. Moreover, these methods generally lack a comprehensive theoretical foundation, often providing neither consistency guarantees nor asymptotic justification for inference. Our proposed ROMA framework unifies these isolated approaches by generalizing traditional linear structural equation models by accommodating object-valued exposures, mediators, and outcomes, where these existing methods can be viewed as special cases. Leveraging Fr\'echet embeddings, ROMA offers a flexible class of models capable of capturing complex, nonlinear dependencies in mediation analysis while supporting rigorous large-sample inference. Furthermore, current causal inference methods for random objects \citep{linCausalInferenceDistribution2023, kurisu_geodesic_2024, bhattacharjee_doubly_2024} can be viewed as special cases of our framework where the mediation pathway is absent.
    
    \item \textbf{An additive RKHS framework for structural causal identification.}  While prior weak conditional mean approaches \citep{li_dimension_2022, sang_nonlinear_2022, bhattacharjeeNonlinearGlobalFrechet2023} are restricted to single-predictor settings, recent extensions into additive Fréchet regression \citep{yang2025variable} focus exclusively on variable selection. These methods lack the structural framework necessary for the decomposition of the outcome into components to separate direct and indirect effects, which is a fundamental challenge in mediation analysis. Our novelty lies in 
    {adopting} an additive RKHS structure that assigns distinct kernel embeddings to predictors, as detailed in Section \ref{sec: Add RKHS}, for solving this structural identification problem.
    Unlike existing methods, our approach bridges weak conditional means with potential outcome theory to rigorously disentangle the natural direct and indirect effects for object-valued data. To our knowledge, this is the first additive RKHS framework designed for causal structural identification. Beyond mediation, this yields a general framework for multivariate object-on-object regression that accommodates multiple complex predictors while preserving their individual geometry and embeddings.
    \item \textbf{A new asymptotics theory for weak conditional mean-type methods.} The weak conditional mean framework was originally introduced for functional sufficient dimension reduction \citep{li_dimension_2022} and later adapted for object-on-object regression \citep{bhattacharjeeNonlinearGlobalFrechet2023,yang2025variable}. However, a formal inferential theory has remained underdeveloped. We provide the global asymptotic normality result (See Theorem \ref{thm: aymp}), moving beyond the pointwise convergence results in \citet{sang_nonlinear_2022}. Notably, our theory achieves a comparable convergence rate but without relying on the variance-dominance condition of \cite{sang_nonlinear_2022} that requires regularization bias to be asymptotically negligible. This contribution extends beyond ROMA, offering new analytical tools for the asymptotic properties of weak conditional mean-based methods. 

    \item \textbf{A rigorous resampling-free inference framework.} Uncertainty quantification in random object analysis often relies on computationally intensive resampling procedures due to the lack of closed-form asymptotics; see, e.g., \cite{Coffman03092023, zhangetldistM} and \cite{bhattacharjee_doubly_2024}. We eliminate the computational burden of resampling by developing global test statistics with tractable asymptotic null distributions in Theorems \ref{thm: TNDE}--\ref{thm: TNIE}. This enables a rigorous resampling-free approach to construct confidence regions and hypothesis tests from asymptotic approximations.
\end{itemize}

The rest of this paper is organized as follows. Section \ref{sec: model const} defines the mediation model for random objects and identifies the causal effects. Section \ref{sec: est} first introduces the additive RKHS framework and the estimation procedure via weak conditional means. Our main theoretical results are provided in Section~\ref{sec: converge}. Section~\ref{sec: simu} evaluates the model’s performance across various data types in simulation studies. Section~\ref{sec: RDA} illustrates the utility of our method in two real-world applications. Section~\ref{sec: conclusion} includes a few concluding remarks. All proofs and additional simulations are presented in the Supplementary Material.

\section{Causal Framework and Identification}\label{sec: model const}

This section establishes the probabilistic foundations of ROMA. Section \ref{sec: MMHE} embeds the outcome space into a Hilbert space and defines the target causal estimands. Section \ref{sec: ICE} specifies the assumptions for identification and derives the identification results. Estimation of these quantities using an additive RKHS framework will be developed in Section \ref{sec: est}.

\subsection{Outcome Embedding and Causal Estimands}\label{sec: MMHE}
Let $(\Omega_X, d_X)$, $(\Omega_M, d_M)$, and $(\Omega_Y, d_Y)$ denote the separable metric spaces associated with the exposure $X$, the mediator $M$, and the outcome $Y$, respectively, all defined on a common probability space $(\Omega, \mathcal{F}, P)$. We adopt the potential outcomes framework to characterize the causal mechanism in mediation analysis. Let $M(x)$ denote the potential mediator value associated with exposure $x \in \Omega_X$, and let $Y(x, m)$ denote the potential outcome associated with exposure $x$ and mediator $m \in \Omega_M$. The observed random elements are related to these potential outcomes via the Stable Unit Treatment Value Assumption (SUTVA) \citep{SUTVA}, such that $M = M(x)$ and $Y = Y(X, M)$. To ensure the identification of causal effects, we require the following standard conditional independence assumptions  (see \cite{song2020bayesian} for a detailed explanation): for all $x, x' \in \Omega_X$ and $m \in \Omega_M$, (i) $Y(x,m)\indep X$, (ii) $Y(x,m)\indep M|X$, (iii) $M(x)\indep X$, and (iv) $Y(x,m)\indep M(x')$.

In traditional mediation analysis, the potential outcome $Y(x,m)$ is typically a real-valued random variable, and causal estimands are defined based on the expectation $\mathbb{E}[Y(x,m)]$. For random objects residing in a general metric space $(\Omega_Y, d_Y)$, this concept generalizes to the Fréchet mean \citep{frechet1948elements}, defined as the minimizer of the expected squared distance:
\begin{equation}\label{eq: Eo}
    \Eo [Y(x,m)]=\argmin_{y\in \Oy} \E[d_Y^2(Y(x,m),y)]
\end{equation}
However, a fundamental challenge arises in defining treatment effects: standard causal contrasts (e.g., the average treatment effect) rely on arithmetic subtraction, an operation that is undefined in metric spaces lacking a linear vector structure. To overcome this limitation, we embed the metric space $\Omega_Y$ into a Hilbert space where linear operations are well-defined.
\begin{assumption}\label{as: rho}
    There is a separable Hilbert space $\c H$ and an isometric map $\rho:\Oy\rightarrow \c H$ such that $\|\rho(y_1) - \rho(y_2)\|_{\mathcal{H}} = d_Y(y_1, y_2)$ for all $y_1, y_2 \in \Omega_Y$.
\end{assumption}

This assumption is satisfied whenever the squared metric $d^2_Y$ is of negative type \citep[Proposition 3]{sejdinovicDistanceRKHS2013}. Formally, a space $(M, d)$ with a semi-metric $d$ is of negative type if for all $n \geq 2, z_1, z_2, \ldots, z_n \in M$ and $\alpha_1, \alpha_2, \ldots, \alpha_n \in \mathbb{R}$, with $\sum_{i=1}^n \alpha_i=0$, one has $\sum_{i=1}^n \sum_{j=1}^n \alpha_i \alpha_j d\left(z_i, z_j\right) \leq 0$. This condition is mild, as the class of negative type spaces includes all separable Hilbert spaces and many other metric spaces encountered in practice.

Throughout this paper, let $V=\rho(Y)$ be the embedded outcome in $\c H$, and
$V(x,m)=\rho(Y(x,m))$ the embedded potential outcome. For a random element $V:\Omega\to\c H$ satisfying $\E\|V\|<\infty$, the expectation $\E[V]\in\c H$ is defined as the Bochner integral {(see, for example, \cite{hsing2015theoretical})}. By the Riesz representation theorem, this is the unique element satisfying $f\mapsto \E\langle f,V\rangle_{\c H}$ for all $f\in\c H$.

While causal effects in Euclidean settings are naturally expressed as differences in expectations (e.g., $\E[Y_1]-\E[Y_2]$), this arithmetic structure is absent in general metric spaces. Although the Fr\'echet mean serves as a valid measure of central tendency, the linear contrasts between Fr\'echet means are not well-defined and lack a canonical interpretation due to the lack of a vector space structure. By embedding outcomes into the Hilbert space $\c H$, the difference $\E[V_1]-\E[V_2]$ is well-defined and can be analyzed using standard linear tools for estimation and inference. To ensure these linear contrasts on the embedded Hilbert space $\c H$ correspond to meaningful objects in the original space, we impose the following assumption:

\begin{assumption}[Fr\'echet Embedding]\label{as: Fre Embd}
    $\E[V(x, m)]\in \rho(\Oy)$ for every $x\in \Omega_X, m\in \Om$.
\end{assumption}

This assumption guarantees that the expected embedded outcome maps back to a valid element in $\Omega_Y$. Sufficient conditions for Assumption \ref{as: Fre Embd} include the convexity and closedness of the image $\rho(\Omega_Y)$ in $\mathcal{H}$ \citep{bhattacharjee_doubly_2024}. A canonical example is the space of square-integrable functions $L^2$, where the embedding is the identity map and the space is itself a Hilbert space. As shown below, several other data types also satisfy this condition.

\begin{example}[Univariate distributional objects]\label{ex: univariate}
  Let $\mathcal{W}_2$ denote the space of absolutely continuous one-dimensional univariate probability distributions, such that for $G\in \c W_2$, $\int_{\R} y^2 \, dG(y) < \infty$. $\mathcal{W}_2$ is equipped with the 2-Wasserstein distance $d_W$. For $G_1, G_2 \in \mathcal{W}_2$, the 2-Wasserstein distance is $ d_W^2(G_1, G_2) = \int_0^1 \left(G_1^{-1}(t) - G_2^{-1}(t)\right)^2 dt,$
    where $G_1^{-1}, G_2^{-1}$ are the quantile functions of $G_1, G_2$ respectively. The Hilbert embedding $\rho: \mathcal{W}_2 \rightarrow \mathbb{L}^2[0,1]$ is defined as $\rho(G)(t) = G^{-1}(t)$. By Theorem 4 of \cite{kolouri2016sliced}, $(\c W_2, \rho)$ is isometric. If the quantile function $G^{-1}(t)\in \rho(\mathcal{W}_2)$ is random, it is easy to verify that $\E[G^{-1}]$ defined by $\E[G^{-1}](t) = \E[G^{-1}(t)]$ for every $t\in[0,1]$ is a quantile function. Therefore, both Assumptions \ref{as: rho}-\ref{as: Fre Embd} hold. 
\end{example}

\begin{example}[Symmetric positive Semi-definite matrices]     Let $\mathcal{S}_p^+$ denote the space of $p\times p$ symmetric positive semi-definite matrices, equipped with the Frobenius metric $d_F(A, B) = \|A - B\|_F$. This space can be embedded into the Hilbert space of symmetric matrices $\mathcal{H} = \mathbb{R}^{p \times p}_{sym}$ equipped with the trace inner product $\langle A, B \rangle_{\mathcal{H}} = \text{tr}(A^\top B)$. The embedding map $\rho: \mathcal{S}_p^+ \rightarrow \mathcal{H}$ is the identity, $\rho(A) = A$, which is trivially isometric. Since $\mathcal{S}_p^+$ forms a closed, convex cone within $\mathcal{H}$, the expectation of any random SPD matrix (with respect to the Frobenius metric) remains within $\mathcal{S}_p^+$. Therefore, the pair $(\mathcal{S}_p^+, \rho)$ satisfies Assumptions \ref{as: rho}--\ref{as: Fre Embd}. 
\end{example}

Under  Assumptions \ref{as: rho} {and} \ref{as: Fre Embd}, Theorem 1 of \cite{bhattacharjee_doubly_2024} yields the  Fréchet Identity ensuring that the Fr\'echet mean $\Eo Y(x,m)$ can be recovered by mapping the Hilbert-space expectation of the embedded random variable $V(x,m)$ back to the original space $\Omega_Y$:
\begin{equation}\label{eq: rho and Eo}
   \text{(Fr\'echet Identity) }\;\Eo Y(x,m)=\rho^{-1} \circ \E [V(x,m)].
\end{equation}
Consequently, causal contrasts can be rigorously formulated within the linear Hilbert space $\c H$ and subsequently interpreted as geometric distances in $\Omega_Y$.

Specifically, we define the Total Effect (TE) as the vector in $\c H$ capturing the
shift in the embedded potential outcome induced by varying the exposure from a fixed reference level $x^*$ to $x$, where $TE:=TE(x, x^*)$ will be used by assuming that $(x, x^*)$ are fixed for simplicity: 
$$TE=\E[V(x,M(x))]-\E[V(x^*,M(x^*))].$$
By exploiting the isometry of $\rho$ and the identity \eqref{eq: rho and Eo}, the magnitude of this vector corresponds directly to the metric distance between the Fréchet means of the potential outcomes:
\begin{align*}
\|TE\|_{\c H} =\left\|\E[V(x,M(x))]-\E[V(x^*,M(x^*))]\right\|_{\c H}
=d_Y\left(\Eo[Y(x,M(x))], \Eo[Y(x^*,M(x^*))]\right).
\end{align*}
Thus, $\|TE\|_{\c H}$ interprets the causal impact as a geometric distance in $\Omega_Y$, generalizing the Euclidean total effect \citep{song2020bayesian} to general metric spaces.

\begin{remark}
 In the absence of a mediator, our TE aligns with the geodesic average treatment effect (GATE) \citep{kurisu_geodesic_2024}, defined as the geodesic connecting the Fr\'echet means $\Eo[Y(x^*)]$ and $\Eo[Y(x)]$.  In the Hilbert space $\c H$, the unique geodesic between $\E[V(x^*)]$ and $\E[V(x)]$ is the line segment characterized by the difference vector $TE = \E[V(x)] - \E[V(x^*)]$. By the isometry of $\rho$ and the Fr\'echet Identity \eqref{eq: rho and Eo}, the inverse image of this segment corresponds to the geodesic in $\Omega_Y$. Thus, our framework includes GATE as a special case. 
\end{remark}

\begin{remark}\label{rem: cod}
    Our framework encompasses the causal effect map for distributional outcome proposed by \cite{linCausalInferenceDistribution2023} as a special case. Let $\mathcal{W}_2(\mathcal{I})$ be the space of univariate distributions on an interval $\mathcal{I} \subset \mathbb{R}$ with finite second moments, equipped with the Wasserstein metric $d_W$. \cite{linCausalInferenceDistribution2023} define the causal effect for a binary treatment $x \in \{0,1\}$ relative to a reference distribution $\lambda$ as $\Delta^\lambda = \mathbb{E}[\mathcal{L}_\lambda Y(1) - \mathcal{L}_\lambda Y(0)]$, where $\mathcal{L}_\lambda Y := Y^{-1} \circ \lambda$ is the optimal transport map. The image of $\mathcal{L}_\lambda$ lies in the weighted Hilbert space $L^2(\mathcal{I}, \lambda)$ equipped with the inner product $\langle f, g \rangle_\lambda = \int_{\mathcal{I}} f g \, d\lambda$.  As shown in the support materials of \cite{linCausalInferenceDistribution2023}, the operator $\mathcal{L}_\lambda$ constitutes a valid embedding satisfying Assumption \ref{as: rho}--\ref{as: Fre Embd}. Consequently, in the absence of a mediator and with reference treatments $(x, x^*) = (1, 0)$, our defined total effect $TE = \mathbb{E}[V(1)] - \mathbb{E}[V(0)]$ is identical to the causal map $\Delta^\lambda$. 
\end{remark}

The Total Effect can be further decomposed into the Natural Direct Effect (NDE) and the Natural Indirect Effect (NIE) within the Hilbert space $\mathcal{H}$:
\begin{align*}
TE &= \underbrace{{\E[V(x,M(x))] - \E[V(x^,M(x))]}}_{\text{NDE}} + \underbrace{{\E[V(x^,M(x))] - \E[V(x^,M(x^))]}}_{\text{NIE}}.
\end{align*}
The NDE captures the shift in the expected embedded outcome attributable solely to the change in exposure $x \to x^*$, while the mediator $M(x)$ remains fixed at its natural value under exposure $x$. Conversely, the NIE isolates the shift attributable to the change in the mediator $M(x) \to M(x^*)$, while holding the exposure fixed at the reference level $x^*$. Our primary objective is to develop a rigorous framework for the estimation and statistical inference of these Hilbert-valued quantities (TE, NDE, and NIE) and map them back to the original metric space to interpret the causal mechanism.

\subsection{Identification of Causal Effects} \label{sec: ICE}
This section details the construction of the RKHS required to identify the proposed causal effects. We begin by defining the kernels for the exposure and mediator spaces.
\begin{assumption}\label{as: Om}
    There exists a continuous positive definite kernel function $\ka_M(\cdot,\,\cdot)$ that generates the separable RKHS $\ca M$ on $\Om$ with inner product $\langle\cdot\,,\cdot\rangle_{\ca M}$.    Similarly,  a continuous positive definite kernel $\ka_X$ on $\Omega_X$ generates the separable RKHS $\c X$ on $\Omega_X$ with $\langle\cdot\,,\cdot\rangle_{\c X}$.
\end{assumption}
A canonical example satisfying Assumption \ref{as: Om} is the Euclidean space $\Omega_X = \R^p$ equipped with the linear kernel $\ka(x, y) = x^Ty+c$ for some $c> 0$.
More generally, valid kernels can be constructed for any metric space $(\Omega_X, d_X)$ of negative type, a condition that is also sufficient for Assumption \ref{as: rho}. Thus, the geometric arguments justifying Assumption \ref{as: rho} analogously guarantee the existence of valid RKHSs for both the exposure and mediator.

\begin{example}[Spherical data]  Let $\mathbb{S}^{p-1} = \{x \in \mathbb{R}^p : \|x\| = 1\}$ denote the unit sphere in $\mathbb{R}^p$. The natural metric on this space is the geodesic distance (great-circle distance), defined as the angle between two points: $d_X(x, y) = \arccos(x^\top y)$. This metric is known to be of negative type \citep{bogomolny2008distance}. Consequently, for any fixed anchor point $x_0 \in \mathbb{S}^{p-1}$, the distance-induced kernel    $ \ka_X(x, y) = \frac{1}{2} [ d_X(x, x_0) + d_X(y, x_0) - d_X(x, y) ]  $
    is positive definite. Since $d_X$ is continuous, the resulting kernel $\ka_X$ is continuous. Given that $\mathbb{S}^{p-1}$ is a separable metric space, Theorem 2.7.5 of \cite{hsing2015theoretical} guarantees that the generated RKHS $\mathcal{X}$ is separable. Thus, Assumption \ref{as: Om} is satisfied. 
\end{example}

Building on the framework of linear mediation models \citep{song2020bayesian}, we impose an additive decomposition on the embedded outcome to separate direct and indirect pathways:
\begin{assumption}\label{as: VIidpVD}
    The embedded outcome $V$ admits a decomposition $V=V_I+V_D$, satisfying the conditional independence conditions that $V_I \indep X|M$ and $V_D \indep M|X$.
\end{assumption}
Under this assumption, $V_D$ captures the (unmediated) direct effect of the exposure $X$, while $V_I$ captures the indirect effect transmitted through the mediator $M$. Crucially, Assumption \ref{as: VIidpVD} implies that the conditional mean of the embedded outcome is additive: $\mathbb{E}[V \mid X, M] = \mathbb{E}[V_D \mid X] + \mathbb{E}[V_I \mid M].$ To formally characterize these components and their dependencies, we proceed by defining the necessary cross-covariance operators and moment conditions; see \cite{bhattacharjeeNonlinearGlobalFrechet2023} for related theoretical foundations.

\begin{assumption}\label{as: mu and sigma}$\E[\ka_M(M,M)] <\infty,\,  \E[\sqrt{\ka_M(M,M)}\|V_I\|] < \infty,$ $ \E[\ka_X(X,X)] <\infty,$ and
$\E[\sqrt{\ka_X(X,X)}\|V_D\|] < \infty, \E[\sqrt{\ka_X(X,X)}\|\tm(M)\|] < \infty$. 
\end{assumption}
Let $\tm$ be the Aronszajn map $m \mapsto \ka_M(\dcd m)$, and $\tau_X(x) = \ka_X(\dcd x)$. Then, for the random element $M$ taking values in $\Om$, $\tm(M)$ is a random element that resides in $\tm(\Om)\subset \c M$. We define the mean elements in the RKHS space $\mu_M=\E[\tm(M)]$ and $\mu_X=\E[\tau_X(X)]$. The covariance operators $\Sigma_{MM}=\E[(\tm(M)-\mu_M)\otimes (\tm(M)-\mu_M)]$ and $\Sigma_{MV}=\E[(\tm(M)-\mu_M)\otimes (V_I-\E V_I)]$ are well defined by Theorem 2.6.5 of \cite{hsing2015theoretical}. Let $\Sigma_{MM}^\dagger=[\Sigma_{MM}|_{\overline{ran(\Sigma_{MM})}}]^{-1}$ be the Moore-Penrose inverse for $\Sigma_{MM}$, {where $ran(\Sigma_{MM})$ is the range of the operator $\Sigma_{MM}$, $\overline{ran(\Sigma_{MM})}$ is the closure of $ran(\Sigma_{MM})$, and $\Sigma_{MM}|_{\overline{ran(\Sigma_{MM})}}$ is the operator $\Sigma_{MM}$ with its domain restricted on $ran(\Sigma_{MM})$. Note that $\Sigma_{MM}|_{\overline{ran(\Sigma_{MM})}}$ is an injection from $\overline{ran(\Sigma_{MM})}$ and ${ran(\Sigma_{MM})}$, so its inverse map is defined. }
Similarly, $\Sigma_{XX}=\E[(\tau_X(X)-\mu_X)\otimes (\tau_X(X)-\mu_X)]$, $\Sigma_{XM}=\E[(\tau_X(X)-\mu_X)\otimes (\tm(M)-\mu_M)]$. $\Sigma_{XV}=\E[(\tau_X(X)-\mu_X)\otimes (V_D-\E V_D)]$, and
$\Sigma_{XX}^\dagger=[\Sigma_{XX}|_{\overline{ran(\Sigma_{XX})}}]^{-1}$.

\begin{assumption}\label{as: rans}
    $\ran(\Sigma_{XM})\subset \ran(\Sigma_{XX})$,
    $\ran(\Sigma_{XV})\subset \ran(\Sigma_{XX})$,
    $\ran(\Sigma_{MX})\subset \ran(\Sigma_{MM})$,
    $\ran(\Sigma_{MV})\subset \ran(\Sigma_{MM})$. The regression operators $R_{XM}=\Sigma_{XX}^\dagger\Sigma_{XM}$, $R_{XV}=\Sigma_{XX}^\dagger\Sigma_{XV}$, $R_{MV}=\Sigma_{MM}^\dagger\Sigma_{MV}$ are bounded operator.
\end{assumption}

Assumption \ref{as: rans} is used to establish the existence of RKHS regression operators \citep{li_dimension_2022,bhattacharjeeNonlinearGlobalFrechet2023}. Under these assumptions, we introduce the weak conditional expectation $\E[V_I\bbar M]$ to characterize the dependence between the embedded variables:
\begin{definition}\label{def: weakcon}
    Let $R_{MV}: \mathcal{H} \to \mathcal{M}$ be the bounded linear operator satisfying $\langle R_{MV}(v), f \rangle_{\mathcal{M}} = \text{Cov}( \langle V_I, v \rangle_{\mathcal{H}}, f(M) )$ for all $v \in \mathcal{H}, f \in \mathcal{M}$. Let $\lambda(\cdot)$ denote the Riesz representation of the functional $v\mapsto R_{MV}(v)(m)$. Then, the weak conditional expectation of $V_I$ given $M$ with respect to $\c M$, denoted by $\E[V_I\bbar M]$, is defined as $\E[V_I\bbar M]=\E V_I+\lambda(M)-\E[\lambda(M)]$.
\end{definition}

This definition relies on the covariance operators. Specifically, let $\Sigma_{MM}$ be the covariance operator on $\mathcal{M}$, and let $\Sigma_{VM}: \mathcal{M} \to \mathcal{H}$ be the cross-covariance operator defined by $\Sigma_{VM} = \E[(V_I - \E V_I) \otimes (\tm(M) - \mu_M)]$, where $\tm(m) = \ka_M(\cdot, m)$ is the Aronszajn map. The operator $R_{MV}$ is formally given by {$R_{MV} = \Sigma_{MM}^\dagger\Sigma_{MV} $}, where $\Sigma_{MM}^\dagger$ denotes the Moore-Penrose pseudoinverse. Its adjoint is given by $R_{VM} = \Sigma_{VM} \Sigma_{MM}^\dagger$. The following proposition provides the explicit closed-form expression for the weak conditional mean.
\begin{proposition}\label{prop: con}
 For any $v\in \c H$, the weak conditional mean $\Psi(m)=\E[V_I\bbar M=m]$ satisfies   
 
     $(1)\;   \langle \Psi(m),v \rangle_{\c H}=\langle \E V_I, v\rangle_{\c H}+R_{MV}(v)(m)-\E[R_{MV}(v)(M)].$
    
     $(2)\;\Psi(m)=\E V_I+R_{VM}\left(\tm(m)-\mu_M\right)=
    \E V_I+\Sigma_{VM}\Sigma_{MM}^\dagger\left(\tm(m)-\mu_M\right)$.
\end{proposition}

The form derived in Proposition \ref{prop: con} (2) reveals that the weak conditional mean is structurally identical to a linear regression model in the feature space. Notably, when the kernels are linear, $\E[V_I \bbar M]$ simplifies to the linear regression.
\begin{proposition}\label{prop: linear weak}
    If $\Om = \R^p$ and $\c H = \R$, and the covariance matrix of $M$ is invertible, and take $\ka_M$ as the linear kernel $\ka_M(m_1, m_2) = c+ m_1^Tm_2$ for some constant $c$. Then $
    \E[V_I\bbar M=m] = \beta_0 + \beta^T (m-\E[M])$, where $\beta_0=\E[V_I]$ and $\beta= [\cov(V_I, M)][\var(M)]^{-1} \in \R^p$.
\end{proposition}
In canonical Linear Structural Equation Models (LSEMs), it is standard to assume that $\E[V_I|M=m] = \beta_0 + \beta^T (m-\E[M])$, that is, $\E[V_I|M=m] =\E[V_I\bbar M=m]$. We assume that this identity holds for all the conditional expectations, generalizing the traditional LSEMs:
\begin{assumption}\label{as: bbar bar} 
The conditional expectations equal their weak counterparts: 
    $\E[\tm(M)| X=x] = \E[\tm(M)\bbar X=x]$, 
    $\E[V_I| M=m] = \E[V_I\bbar M=m]$,  and $\E[V_D| X=x] = \E[V_D\bbar X=x]$.
\end{assumption}
This assumption is naturally satisfied when the chosen kernels $\ka_M$ and $\ka_X$ are universal, i.e., when their induced RKHSs are dense in the space of continuous real-valued functions. Canonical examples of universal kernels on complete separable metric spaces include the Gaussian and Laplacian kernels \citep{li_dimension_2022,zhang2024NonlinearSDR}.

Let $\Phi(x)=\E[\tm(M)\bbar X=x]$, and $\bgamma(x)=\E[V_D\bbar X=x]$. 
Synthesizing the additive decomposition (Assumption \ref{as: VIidpVD}) with the model specification (Assumption \ref{as: bbar bar}) yields the following structural equations that govern our framework: $\E[\tm(M)| X=x]=\Phi(x)$ and $\E[V| X=x,M=m]= \E[V_I| M=m] +\E[V_D| X=x]=\Psi(m)+\bgamma(x).$

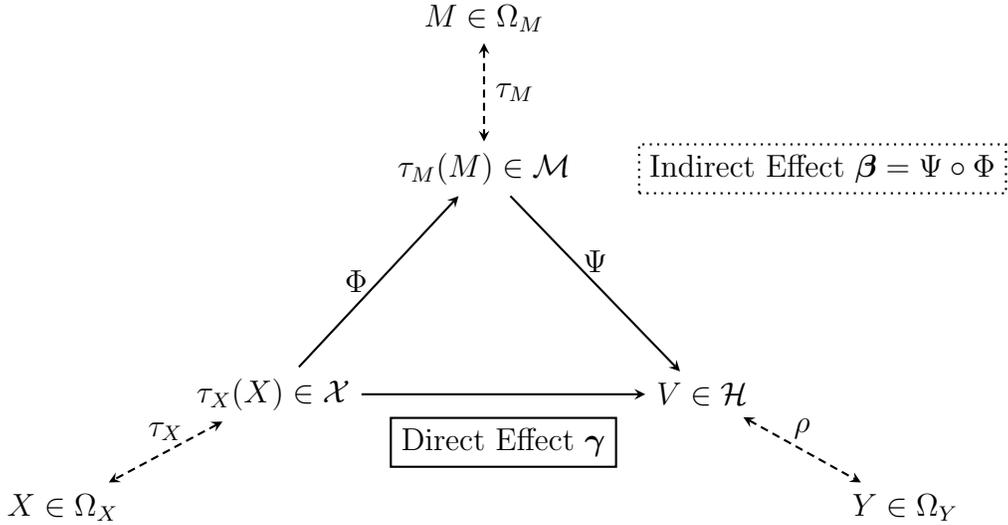
\begin{figure}[htp]
\centering
\begin{tikzpicture}[->, >=stealth, node distance=2.5cm, thick]
    \node (x2)  {$\tau_X(X) \in \c X$};
    \node (x3) [left of=x2, xshift=-0.3cm,yshift=-1.5cm]{$X \in \Omega_X$};
    \node (m2) [right of=x2, xshift=0.3cm, yshift=3cm] {$\tm(M) \in \c M$};
    \node (m1) [right of=x2, xshift=0.3cm, yshift=5cm] {$M \in \Om$};
    \node (y2) [right of=x2, xshift=3.2cm] {$V \in \c H$};
    \node (y3) [right of=y2, xshift=0.2cm, yshift=-1.5cm] {$Y \in \Oy$};
    \node (ind) [right of=m2, xshift=2cm,draw,dotted] {Indirect Effect $\bbeta=\Psi\circ \Phi$};

    \draw[->] (x2) -- node[below, yshift=-0.3cm, draw] {Direct Effect $\bgamma$} (y2);
    \draw[->,] (x2) -- node[left] {$\Phi $} (m2);
    \draw[->] (m2) -- node[above] {$\Psi$} (y2);
    \draw[<->, densely dashed] (y2) --node[above]{$\rho$} (y3);
    \draw[<->, densely dashed] (x2) --node[above]{$\tau_X$} (x3);
    \draw[<->, densely dashed] (m1)--node[right]{$\tm$}(m2);
\end{tikzpicture}
\caption{Path diagram illustrating the causal structure of ROMA. Solid arrows denote structural causal relationships, while dashed arrows represent embeddings into Hilbert spaces. The terms $\tau_X$ and $\tau_M$ denote the Aronszajn feature maps defined by $x \mapsto \ka_X(\cdot, x)$ and $m \mapsto \ka_M(\cdot, m)$, respectively, while $\rho$ represents the isometric outcome embedding. The operators $(\Phi, \Psi, \bgamma)$ represent weak conditional mean operators.
}
\label{fig: diag}
\end{figure}

The causal structure of our framework is illustrated in Figure \ref{fig: diag}. Recall from Proposition \ref{prop: con}(2) that $\Psi$ depends on $m$ strictly through the feature map $\tm(m)=\ka_M(\dcd m)\in \c M$. To facilitate operator composition, we slightly abuse notation by extending the domain of $\Psi$ from the data space $\Omega_M$ to a subspace of the RKHS $\mathcal{M}$. Specifically, for any $f_M-\mu_M\in \overline{\ran }(\Sigma_{MM})$, we denote $
\Psi\circ f_M=\E V_I + R_{VM}(f_M-\mu_M).$ Hence, $\Psi$ can be viewed as a linear operator on a subspace of $\c M$, and $\Psi\circ \tm(M)$ is exactly the weak conditional mean for $V_I$ given $M$ by Proposition \ref{prop: con}(2). Since we assume that $\ran(\Sigma_{MX})\subset \ran(\Sigma_{MM})$ in Assumption \ref{as: rans}, we have $\Phi(x)-\mu_M \in \ran(\Sigma_{MM})$ for any $x$. Therefore, the composite operator $\Psi\circ \Phi$ is mathematically well-defined, and we denote it by $\bbeta$. With these components formally established, we are now positioned to characterize the target causal effects using the mappings in Figure \ref{fig: diag}.
\begin{lemma}\label{lem: char4NEs}
   Under Assumptions \ref{as: rho}--\ref{as: bbar bar}, we have $ \NDE = \bgamma(x)-\bgamma(x^*)$, and $\NIE = \bbeta(x)-\bbeta(x^*)$.
\end{lemma}

Lemma \ref{lem: char4NEs} shows that the causal effects are characterized by the operators $\bbeta$ and $\bgamma$. Specifically, the NDE is driven by the shift in the direct pathway operator $\bgamma$, while the NIE is driven by the shift in the composite indirect pathway operator $\bbeta$. Consequently, the total effect is simply their sum: $\operatorname{TE}=\bbeta(x)-\bbeta(x^*)+\bgamma(x)-\bgamma(x^*).$
\begin{remark}
Classical LSEMs \citep{song2020bayesian} are a special case of our framework when $\ka_X$ and $\ka_M$ are linear kernels. By Proposition \ref{prop: linear weak}, $\Phi$ and $\Psi$ reduce to linear maps for $\Om=\R^{p_m}$ and $ \Oy=\R$. Suppose they can be expressed as $\Phi(x)=\bbeta_ax, \bbeta_a \in \R^{p_m\times q}$ and $\Psi(m)=\bbeta_mm, \bbeta_m\in \R^{1\times p_m}$. Analogously, $\bgamma\in \R^{1\times q}$ reduces to the linear operator. 
By Lemma \ref{lem: char4NEs}, $\NDE=\bgamma(x-x^*)$ and $\NIE=\bbeta_m\bbeta_a(x-x^*)$, matching those identified in \cite{song2020bayesian}.
\end{remark}
\begin{remark}
    Our model also includes the mediation framework for density-valued mediators \citep{zhangetldistM} as a special example. Their approach assumes $\Omega_X, \Omega_Y \subset \mathbb{R}$ and models the conditional expectations of the mediator's quantile function $M^{-1}(t)$ and the outcome $Y$ as linear functionals. This formulation is structurally identical to our additive RKHS model if we apply a linear kernel for $X$ and the $L^2$ inner product kernel $\ka_M(m_1, m_2) = \int_0^1 m_1^{-1}(t)m_2^{-1}(t)dt$ for $M$, yielding equivalent NDE and NIE. 
\end{remark}

\section{Estimation}\label{sec: est}
This section {develops} the estimation procedures for the ROMA framework. Leveraging the additive RKHS structure of our model, we propose empirical estimators for the structural operators and derive closed-form expressions for the causal effects via coordinate mappings.

\subsection{Estimation within the Additive RKHS Framework} \label{sec: Add RKHS}

A fundamental requirement of mediation analysis is the ability to decompose the outcome into distinct components attributable to the exposure $X$ (NDE) and the mediator $M$ (NIE). In multivariate linear regression, the model $y = \beta_X x + \beta_M m$ inherently separates the additive contributions of $x$ and $m$. However, existing weak conditional mean frameworks \citep{li_dimension_2022, sang_nonlinear_2022, bhattacharjeeNonlinearGlobalFrechet2023} are restricted to a single predictor equipped with a single kernel. This prevents the essential decomposition required for mediation when the exposure and mediator reside in distinct metric spaces with different kernel embeddings.

To overcome this limitation, we develop an additive RKHS framework that generalizes the structural additivity of linear models to general metric spaces. By defining a joint RKHS as the direct sum of the individual predictor spaces, we enable the simultaneous and distinct estimation of the mediator operator $\Psi$ (the effect of $M$) and $\bgamma$ (the effect of $X$). Following \cite{aronszajn1950theory}, we utilize the sum of reproducing kernels to formally construct this joint space, laying the algebraic groundwork for our empirical estimators:
\begin{proposition}\label{prop: KZ}
 Let $K_i:\Omega_i\times \Omega_i \rightarrow \R$ be nonnegative kernels generating RKHS $\mathcal{H}_i$ that has norms $\|\cdot\|_i, i=1,2$. Then, we denote $K$ as a function $ (\Omega_1,\Omega_2)\times (\Omega_1,\Omega_2)\rightarrow \R$, which maps $(x_1,y_1)\times (x_2,y_2)$ to $K_1(x_1,x_2)+K_2(y_1,y_2)$.
Then, $K$ is the  reproducing kernel of $\c H_+$, the set of all functions of the form $f_1(x)+f_2(y):x\in \Omega_1, y\in \Omega_2$ with $f_i \in \mathcal{H}_i, i=1,2$ under the norm $
\|f\|^2=\min _{f_i \in \c H_i, i=1,2: f=f_1+f_2}\left\{\left\|f_1\right\|_1^2+\left\|f_2\right\|_2^2\right\}$.
\end{proposition}

In our setting, we let $(K_1,\Omega_1, \c H_1)$ be $(\ka_X, \Omega_X, \c X)$ and $(K_2, \Omega_2, \c H_2)$ be $(\ka_M, \Om, \c M)$ in Proposition \ref{prop: KZ}. This yields the additive kernel $\ka_Z = \ka_X+\ka_M$, on the product space $\Omega_Z = \Omega_X \times \Omega_M$, together with the corresponding RKHS $\c Z$. Let $Z = (X, M) \in \Omega_X \times \Om$, and the mean embedding in $\mathcal{Z}$ is defined as $\mu_Z = \mu_X+\mu_M$. Let  $\Sigma_{ZZ}=\E[(\ka_Z(\dcd Z)-\mu_Z)\otimes (\ka_Z(\dcd Z)-\mu_Z)]$, $\Sigma_{ZV}=\E[(\ka_Z(\dcd Z)-\mu_Z)\otimes (V-\E V)]$, and $\Sigma_{ZZ}^\dagger=[\Sigma_{ZZ}|_{\overline{ran(\Sigma_{ZZ})}}]^{-1}$. Similar to Assumptions \ref{as: mu and sigma} {and} \ref{as: rans}, $\Sigzv,\Sigz$ and $R_{ZV}$ are {well defined} under Assumption \ref{as: Z basic}:

\begin{assumption}\label{as: Z basic}
    $\E[\ka_Z(Z,Z)] <\infty, \, \E[\sqrt{\ka_Z(Z,Z)}\|V\|] < \infty$. We have $\ran(\Sigma_{ZV})\subset \ran(\Sigma_{ZZ})$, and the centered regression operator $R_{ZV}=\Sigma_{ZZ}^\dagger\Sigma_{ZV}$ is a bounded operator.
\end{assumption}

To guarantee that the additive decomposition is identifiable, Assumption \ref{as: WIWD} serves as an extension of Assumption \ref{as: VIidpVD}. This is a mild structural requirement, analogous to the assumption in linear models that population-level residuals are independent of other covariates. 

\begin{assumption}\label{as: WIWD}
    The population-level residual $V_I-\Psi(M)$ is independent of the exposure $X$, and the residual $V_D-\bgamma(X)$ is independent of the mediator $M$.
\end{assumption}

\begin{lemma}\label{lem: RZVwR}
    If Assumption \ref{as: Om}-\ref{as: WIWD} holds, then for any $f_Z=f_X+f_M\in \c Z$ such that $f_X\in \ran(\Sigx), f_M\in \ran(\Sigm)$, we have $\Sigma_{VZ}\Sigma_{ZZ}^\dagger f_Z= \Sigma_{VX}\Sigma_{XX}^\dagger f_X+ \Sigma_{VM}\Sigma_{MM}^\dagger f_M$.
\end{lemma}
The power of this joint construction is formalized in the above lemma. Just as a multivariate linear model isolates the partial effects of each predictor, Lemma \ref{lem: RZVwR} guarantees that the regression operator on the joint space $\mathcal{Z}$ decomposes into the sum of the individual operators on $\mathcal{X}$ and $\mathcal{M}$. This ensures that no structural interference or mathematical artifacts are introduced by the joint modeling, and the joint operator perfectly preserves the individual structural mappings. This decomposition allows us to express the weak conditional mean in a transparent, additive form. For any joint observation $z=(x,m)\in \Omega_X\times \Om$, we have
\begin{align}
    \E[V\bbar Z=z]=&\E[V]+ \Sigma_{VZ}\Sigma_{ZZ}^\dagger\left(\ka_Z(\dcd z)-\mu_Z\right)\nonumber\\
    =&\E[V]+ \Sigma_{VZ}\Sigma_{ZZ}^\dagger\left(\tm(m)-\mu_M\right)
    + \Sigma_{VZ}\Sigma_{ZZ}^\dagger\left(\tau_X(x)-\mu_X\right)\nonumber\\
    =&\E[V]+ \Sigma_{VM}\Sigma_{MM}^\dagger\left(\tm(m)-\mu_M\right)
    + \Sigma_{VX}\Sigma_{XX}^\dagger\left(\tau_X(x)-\mu_X\right)\nonumber\\
    =& \Psi(m)+\bgamma(x),\label{eq: Vz}
\end{align}
where the second equality follows from the additive construction $\ka_Z=\ka_X+\ka_M$, the third is implied by Lemma \ref{lem: RZVwR}, and the last equality uses Proposition \ref{prop: con}.  This result is the cornerstone of ROMA, as it reduces the estimation of complex, non-linear causal pathways to the tractable task of computing a single weak conditional mean on the joint space $\mathcal{Z}$.

\subsection{Closed-Form Estimation via Coordinate Mappings}\label{sec: CM}
We now derive closed-form estimators for $\Phi, \Psi$, and $\bgamma$ via coordinate mappings, reducing the empirical operator estimation to standard kernel Gram matrix operations.

Before detailing the empirical estimators, we establish our coordinate notation. Let $\E_n[\cdot]$ denote the empirical mean. For a finite-dimensional subspace $\mathcal{L}$ with basis $\mathcal{B} = \{\xi_1, \dots, \xi_p\}$, any element $\xi \in \mathcal{L}$ uniquely maps to a coordinate vector $[\xi]_{\mathcal{B}} = (a_1, \dots, a_p)^\top \in \mathbb{R}^p$ such that $\xi = \sum_{i=1}^p a_i \xi_i$. For a linear operator $A: \mathcal{L}_1 \to \mathcal{L}_2$ between spaces with bases $\mathcal{B}$ and $\mathcal{C}$, we let ${_{\mathcal{C}}}[A]_{\mathcal{B}}$ denote its matrix representation defined by the coordinate mapping $[A \xi]_{\mathcal{C}} = {_{\mathcal{C}}}[A]_{\mathcal{B}} [\xi]_{\mathcal{B}}$.

For $i = 1,\dots, n$, let $(X_i, M_i, Y_i)$ represent the observed exposure, mediator, and outcome, respectively. We denote $\hat\mu_X=\frac{1}{n}\sum_{i=1}^n\ka_X(\dcd X_i), \hat\mu_M=\frac{1}{n}\sum_{i=1}^n\ka_M(\dcd M_i)$, and $\hat \mu_Z = \hat \mu_X+\hat \mu_M$. Suppose the subspace $\overline{\operatorname{ran}}(\hat\Sigma_{XX})$ is spanned by $\mathcal{B}_X=\left\{\ka_X(\dcd X_i)-\hat\mu_X: i=1, \ldots, n\right\}$, equipped with inner product $\langle f,g\rangle_{\c X}=[f]^TK_X[g]$, where $K_X$ is the Gram matrix such that $(K_X)_{ij}=\ka_X(X_i,X_j)$. Likewise, the space $\overline{\ran}(\hat\Sigma_{MM})$ spanned by $\c B_M = \{\ka_M(\dcd M_i)-\hat\mu_M, i=1,\dots n\}$ is equipped with inner product $\langle f,g\rangle_{\c M}=[f]^TK_M[g]$, where $K_M$ is the Gram matrix of $\ka_M$.  We also have $\c B_Z = \{\ka_Z(\dcd Z_i)-\hat\mu_Z, i=1,\dots n\}$ that forms a basis for $\overline{\ran}(\hat\Sigma_{ZZ})$, equipped with inner product with Gram matrix $K_Z=K_M+K_X$.

We estimate $\Sigma_{XX}$ by  $\hat\Sigma_{XX}=\frac 1 n \sum_{i=1}^n(\ka_X(\dcd X_i)-\hat\mu_X)\otimes (\ka_X(\dcd X_i)-\hat\mu_X)$, and $\Sigma_{MX}$ by $\hat{\Sigma}_{MX}=\frac{1}{n}\sum_{i=1}^n (\ka_M(\dcd M_i)-\hat\mu_M)\otimes(\ka_X(\dcd X_i)-\hat\mu_X)$. Let $G_X=Q \ka_X Q$ and $G_X^{\dagger}=(G_X+\epsilon I_n)^{-1}$ be the Moore-Penrose inverse via the Tikhonov regularization to prevent overfitting, where $\epsilon>0$ is a tuning constant and {$Q$ is the projection matrix $I_n-\frac{1}{n} 1_n 1_n^T$.}
We then have the coordinate representations (see \cite{li2018sufficient} for details): 
$
_{\mathcal{B}_X}[\hSigx]_{\mathcal{B}_X}=n^{-1} G_X$, $
{}_{\mathcal{B}_X}[\hat{\Sigma}_{XX}^{\dagger}]_{\mathcal{B}_X}=n G_X^{\dagger}$, and $
{}_{\c B_M}[\hat \Sigma_{MX}]_{\mathcal{B}_X}=n^{-1} G_X.$ Let $c_x = [\ka_X(\dcd X)-\hat\mu_X]_{\c B_X}$. To solve $c_x$, we let $d_x=((d_x)_i)_{i=1,\dots n}$ with $(d_x)_i=\ka_X(x, X_i)-\hat\mu_X(X_i)$. Note that
$(d_x)_i=\langle \ka_X(\dcd x)-\hat \mu_X$, $ \ka_X(\dcd X_i)\rangle_{\c X}=e_i^T \ka_XQ c_x$
implies $d_x=\ka_X Qc_x$. Hence, we obtain $c_x=Q G_X^\dagger d_x$. Plugging the coordinates into Proposition \ref{prop: con}, we have $[\hat\Phi(x)-\hat\mu_M]_{\c B_M}= G_XG_X^{\dagger}c_x$, or equivalently, 
$
\hat\Phi(x)=\frac{1}{n}h_M^T\mathbf{1}_n + h_M^T G_X(G_X+\epsilon I_n)^{-1}c_x,
$
where $h_M=(\tm(M_1),\dots \tm(M_n))^T.$

Next, we estimate $\Psi, \bgamma$ via coordinate mappings. Let $\hSigz$ and $\hat{\Sigma}_{VZ}$ be the estimates for $\Sigma_{ZZ}$ and $\Sigma_{VZ}$, respectively. Suppose the space $\c H$ is spanned by $\mathcal{C}_V=\left\{V_i-\E_n[V]: i=1, \ldots, n\right\}$. Let $G_Z=Q K_Z Q$ and $G_Z^{\dagger}=\left(G_Z+\teps I_n\right)^{-1}$ with $\teps>0$. Let $\tilde c_z, \tilde c_x,\tilde c_m, d_z, d_m$ be the $n$-dimensional  vectors such that  $
    \tilde c_z=[\ka_Z(\dcd z)-\hat\mu_Z]_{\c B_Z}$, $\tilde c_x=[\ka_X(\dcd X)-\hat\mu_X]_{\c B_Z}, \tilde c_m=[\ka_M(\dcd M)-\hat\mu_M]_{\c B_Z}$, and for $i=1,\dots n$, $
    (d_m)_i=\ka_M(m,M_i)-\hat\mu_M(M_i)$, $(d_z)_i=\ka_Z(z,Z_i)-\hat\mu_Z(Z_i)$.
Note that $\tilde c_x$ differs from $c_x$:  $\tilde c_x$ is the coordinate representation of $\ka_X(\dcd X)-\hat\mu_X$ in the joint basis $\c B_Z$, while $c_x$ is the coordinate representation in the basis $\c B_X$.

To derive the explicit coordinate representations, we establish a fundamental property of the additive RKHS defined in Proposition \ref{prop: KZ}. Specifically, the component spaces $\c X$ and $\c M$ are combined orthogonally, while preserving their internal inner products and norms. 
\begin{lemma}\label{lem: accord}
    Let $\langle\cdot \,,\, \cdot\rangle_+$ and  $\|\cdot \|_+$ be the inner product and the norm on the joint space $\c H_+$. For $f_1, f_2$ both belong to $\c H_1$ (or respectively in $\c H_2$), the inner products and norms are preserved: $\langle f_1,f_2\rangle_{\c H_1} = \langle f_1,f_2\rangle_+$ and $\|f_1\|_1=\|f_1\|_+$ (or $\langle f_1,f_2\rangle_{\c H_2} = \langle f_1,f_2\rangle_+$ and $\|f_1\|_2=\|f_1\|_+$). Moreover, $\langle f, g\rangle_+=0$ for any $f\in \c H_1$ and $g\in \c H_2$.
\end{lemma}
By applying Lemma \ref{lem: accord}, for $i=1,\dots, n$, we have $(d_x)_i=\langle \ka_X(\dcd x)-\hat\mu_X,\ka_X(\dcd X_i)\rangle_{\c Z}=\langle \ka_X(\dcd x)-\hat\mu_X$, $\ka_Z(\dcd Z_i)\rangle_{\c Z}=e_i^T K_ZQ \tilde c_x$, and
$(d_m)_i=\langle \ka_M(\dcd m)-\hat\mu_M,\ka_M(\dcd M_i)\rangle_{\c Z}=\langle \ka_M(\dcd m)-\hat\mu_M$, $\ka_Z(\dcd Z_i)\rangle_{\c Z}=e_i^T K_ZQ \tilde c_m$. Solving these linear systems yields $\tilde c_x=Q G_Z^\dagger d_x, \tilde c_m=Q G_Z^\dagger d_m$ and $\tilde c_z=Q G_Z^\dagger d_z=\tilde c_x+\tilde c_m$. Let $h_V=(V_1,\dots, V_n)^T$, and substituting these coordinate representations into \eqref{eq: Vz}, we obtain the closed-form estimator:
\begin{align*}
    \hat{\Psi}(m)+\hat{\bgamma}(x)
    =\frac{1}{n}h_V^T\mathbf{1}_n + h_V^T G_Z (G_Z+\teps I_n)^{-1} \tilde c_m+ h_V^T G_Z (G_Z+\teps I_n)^{-1}\tilde c_x.
\end{align*}

Recall that the indirect pathway operator is defined by the composition  $\bbeta=\Psi\circ\Phi$. To compute the empirical counterpart $\hat\bbeta = \hat\Psi \circ \hat \Phi$, we utilize the coordinate representation $[\hat\Phi(x)-\hat\mu_M]_{\c B_M} = G_XG_X^{\dagger}c_x$. Substituting it with $\tilde c_m$ into the expression of $\hat\Psi(m)$, we have
\begin{align*}
    \hat\bbeta(x)+\hat\bgamma(x)=\frac{1}{n}h_V^T\mathbf{1}_n + h_V^T G_Z (G_Z+\teps I_n)^{-1} G_XG_X^{\dagger}c_x+ h_V^T G_Z (G_Z+\teps I_n)^{-1}\tilde c_x.
\end{align*}

Building on these explicit expressions, it is straightforward to estimate the causal effects $\NDE$ and $\NIE$ established in Lemma \ref{lem: char4NEs}. We substitute the operator definitions to obtain $\NDE = \Sigma_{VZ}\Sigma_{ZZ}^\dagger\left[\tau_X(x)-\tau_X(x^*)\right]$, and 
$\NIE = \Sigma_{VZ}\Sigma_{ZZ}^\dagger\Sigma_{MX}\Sigma_{XX}^\dagger\left[\tau_X(x)-\tau_X(x^*)\right]$.
Thus, denoting $\tilde c_x$ (or $\tilde c_{x^*}$) as the coordinate representation of $\tau_X(x)-\hat\mu_X$ (or $\tau_X(x^*)-\hat\mu_X$) with respect to the joint basis $\c B_Z$, we obtain the closed-form estimator for the NDE as follows:
\begin{align*}
\hNDE
=\hat\Sigma_{VZ}\hat\Sigma_{ZZ}^\dagger\left[\tau_X(x)-\tau_X(x^*)\right] =h_VG_Z(G_Z+\teps I_n)^{-1}(\tilde c_x-\tilde c_{x^*}).
\end{align*}
Let $c_x$ (or $c_{x^*}$) be the coordinate representation of $\tau_X(x)-\hat\mu_X$ (or $\tau_X(x^*)-\hat\mu_X$) with respect to the  basis $\c B_X$. We also have the closed-form estimator for the NIE as follows:
\begin{align*}
    \hNIE
    =\hat\Sigma_{VZ}\hat\Sigma_{ZZ}^\dagger\hat\Sigma_{MX}\hat\Sigma_{XX}^\dagger\left[\tau_X(x)-\tau_X(x^*)\right],
    =h_VG_Z(G_Z+\teps I_n)^{-1}G_X(G_X+\epsilon I_n)^{-1}(c_x-c_{x^*}).
\end{align*}

In practice, we suggest using the generalized cross-validation (GCV) score to choose the parameters $(\epsilon, \teps)$ in the Tikhonov-regularization. In fitting $\hat\Phi$, we consider
$$
\operatorname{GCV}(\epsilon) = \frac{1}{n}\sum_{i=1}^n\frac{\|M_i-\hat\Phi(X_i)\|^2}{\left(1-\tr[QG_X(G_X+\epsilon I_n)^{-1} + 1_n 1_n^T / n]/n\right)^2}.
$$
The numerator of this criterion measures the prediction error, while the denominator penalizes overfitting. Similarly for $\hat \Psi+\hat\bgamma$, we consider
$$
\operatorname{GCV}(\teps) = \frac{1}{n}\sum_{i=1}^n\frac{\|V_i-\hat\Psi(M_i)-\hat\bgamma(X_i)\|^2}{\left(1-\tr[QG_Z(G_Z+\teps I_n)^{-1} + 1_n 1_n^T / n]/n\right)^2}.
$$
We choose the optimal $(\epsilon, \teps)$ via a grid search to minimize the two $\operatorname{GCV}$ scores simultaneously.
\section{Asymptotic Properties}\label{sec: converge}
\subsection{Convergence Rates}

The convergence rates for our empirical operators can be established under further smoothness conditions. Assumption \ref{as: S} characterizes this smoothness, which is a standard condition in RKHS-based functional regression \citep{bing2017SDR,li2018sufficient,bhattacharjeeNonlinearGlobalFrechet2023}.

\begin{assumption}\label{as: S}
    There exists bounded linear operators $S_{ZV}: \mathcal{H} \to \mathcal{Z}$ and $S_{XM}: \mathcal{X} \to \mathcal{M}$, along with $\eta,\teta\in (0,1]$, such that $\Sigzv=\Sigz^{1+\teta}S_{ZV}$ and $\Sigma_{XM}=\Sigma_{XX}^{1+\eta}S_{XM}$  are bounded.
\end{assumption}

Let $\|\cdot\|_{OP}$ denote the operator norm  (see \cite{hsing2015theoretical} for the definition). We write $a_n \succ b_n$ if $a_n/b_n \rightarrow 0$; $a_n \prec b_n$ if $b_n/a_n \rightarrow 0$; $a_n \succeq b_n$ if $a_n/b_n \rightarrow L$ for some $L < \infty$; $a_n \preceq b_n$ if $b_n/a_n \rightarrow L$ for some $L < \infty$; and $a_n \asymp b_n$ if $a_n/b_n \rightarrow 1$.
 
\begin{proposition}\label{prop: R error}
    Under Assumptions \ref{as: rho}, \ref{as: Om}, \ref{as: Z basic},
    and \ref{as: S}, we have
    $\|\hat \Sigma_{VZ}\hat \Sigma_{ZZ}^\dagger -\Sigma_{VZ} \Sigma_{ZZ}^\dagger  \|_{OP}=O_P(\teps^{-1} n^{-1/2}+\teps^{\teta})$, and 
   $ \|\hat \Sigma_{XM}\hat \Sigma_{XX}^\dagger -\Sigma_{XM} \Sigma_{XX}^\dagger  \|_{OP}=O_P(\epsilon^{-1} n^{-1/2}+\epsilon^{\eta})$.  
\end{proposition}
Proposition \ref{prop: R error} captures the bias-variance trade-off in Tikhonov-regularized inverse problems. Provided $n^{-1/2} \succ \epsilon \succ 1$ and $n^{-1/2} \succ \teps \succ 1$, the empirical regression operators are statistically consistent. This operator-level convergence naturally guarantees the asymptotic consistency of the predicted conditional means and the estimated causal effects. 

\begin{theorem}\label{thm: ce}
    Under Assumptions \ref{as: rho}--\ref{as: S}, for any $x\in \Omega_X, m\in \Omega_M$, if $\epsilon\prec 1$ and $\teps\prec 1$, then
   $ \|\hat\Phi(x)-\Phi(x)\|=O_P(\epsilon^{-1} n^{-1/2}+\epsilon^{\eta})$,
   $ \|\hat\Psi(m)+\hat\bgamma(x)-\Psi(m)-\bgamma(x)\|=O_P(\teps^{-1} n^{-1/2}+\teps^{\teta})$,
   $ \|\hNDE-\NDE\|=O_P(\teps^{-1} n^{-1/2}+\teps^{\teta})$, and 
   $ \|\hNIE-\NIE\|=O_P(\teps^{-1} n^{-1/2}+\teps^{\teta}+\epsilon^{-1} n^{-1/2}+\epsilon^{\eta})$.
\end{theorem}

\subsection{Asymptotic Normality}
We define the population-level residuals for the mediator and outcome models as $U = \tau_M(M) - \E[\tau_M(M) \mid X]$ and $W = V - \E[V \mid Z]$, respectively. We impose the following standard conditions regarding their independence and finite moments: 
\begin{assumption}\label{as: U}
$X\indep U, (X,M,U)\indep W$. $\E[\|U\|^2]<\infty, \E[\|W\|^2]<\infty$.
\end{assumption}

Let $\{(\lambda_j, \varphi_j)\}_{j=1}^\infty$ denote the eigenvalue-eigenfunction pairs of the marginal covariance operator $\Sigma_{XX}$, sorted in descending order such that $\lambda_1 \ge \lambda_2 \ge \dots > 0$. Similarly, let $\{(\tilde{\lambda}_j, \tilde{\varphi}_j)\}_{j=1}^\infty$ denote the spectral decomposition for the joint covariance operator $\Sigma_{ZZ}$.  
\begin{assumption}\label{as: lambda}$\lambda_j \asymp  j^{-\alpha}$ for some $\alpha>1$, and $\tilde\lambda_j \asymp  j^{-\talpha}$ for some $\talpha>1$.
\end{assumption}
This assumption characterizes the effective capacity of the respective  RKHSs and ensures that the variation of the random elements $X$ and $Z$ is primarily concentrated in the leading, low-frequency spectral components of their covariance operators. 

To quantify the limiting distribution of our estimators, we introduce two functionals for any arbitrary elements $f\in \c Z$ and $g\in \c X$:
\begin{equation}\label{eq: varthetas}
\begin{aligned}
        \vartheta^Z_n[f]=\sum_{j\in \mathbb N}(\tilde\lambda_j+\teps)^{-2}\tilde\lambda_j[\langle \tilde\varphi_j, f\rangle]^2,\quad \text{and} \quad
    \vartheta^X_n[g]=\sum_{j\in \mathbb N}(\lambda_j+\epsilon)^{-2}\lambda_j[\langle \varphi_j,g\rangle]^2.
\end{aligned}
\end{equation}
These functionals capture the interaction between the Tikhonov regularization parameters and the operator spectrum, characterizing the asymptotic variance in the following theorem.

\begin{theorem}\label{thm: aymp}
    Suppose the conditions of Theorem \ref{thm: ce} and Assumptions \ref{as: U}--\ref{as: lambda} hold, and assume further that the reproducing kernel $\ka_X$ is uniformly bounded. For every $x_0\in \Omega_X$, if $\epsilon \succeq n^{-1/(\eta+\frac{3\alpha+1}{2\alpha})}$, $\epsilon \succ n^{-1/2}$, and $n^{-1/2}\{\vartheta_n^X[\tau_X(x_0)]\}^{1/2}\succ \epsilon^{\eta}$, 
    then we have$$\sqrt{n}\vartheta_n^X[\tau_X(x_0)]^{-1/2}\left(\hat\Phi(x_0)-\Phi(x_0)\right)\overset{\c D}
    {\rightarrow} \c G_U,
    $$
    where $\c G_U$ is a mean-zero Gaussian random element with covariance operator $\E[U\otimes U]$. 
\end{theorem}
Here, $\mathcal{G}_U \equiv \mathcal{N}(0, \E[U \otimes U])$ denotes a Gaussian random element taking values in the mediator RKHS $\mathcal{M}$. It is characterized by its continuous linear projections: for any $\zeta \in \c M$,  $\langle \c G_U,\zeta \rangle\sim \c N(0,\E[\langle U, \zeta\rangle^2])$. Using the same strategy developed in Theorem \ref{thm: aymp}, we can straightforwardly establish analogous asymptotic normality results for the structural estimators $\hat{\Psi}$ and $\hat{\bgamma}$, and by extension, for the estimated natural direct and indirect effects.

\begin{corollary}\label{thm: aympZ}
    Suppose the conditions of Theorem \ref{thm: ce} and Assumptions \ref{as: U}--\ref{as: lambda} hold, and assume further that the reproducing kernel $\ka_Z$ is bounded. For any $v \in \c H $, $x_0\in \Omega_X$, and $m_0\in \Omega_M$,
     if $\teps \succ n^{-1/(\teta+\frac{3\talpha+1}{2\talpha})}$, $\teps \succ n^{-1/2}$,  and $n^{-1/2}\left\{\vartheta^Z_n\left[\tau_X(x_0)+\tm(m_0)\right]\right\}^{1/2}\succ \teps^{\teta}$, we have
    $$
    \sqrt{n}\left\{\vartheta^Z_n\left[\tau_X(x_0)+\tm(m_0)\right]\right\}^{-1/2}\left(\hat\bgamma(x_0)-\bgamma(x_0)+\hat\Psi(m_0)-\Psi(m_0)\right)\overset{\c D}
    {\rightarrow} \c N(0,\E[W\otimes W]) .
    $$
\end{corollary}

\begin{theorem}\label{thm: aymcr}
    Suppose the conditions of Theorem \ref{thm: ce} and Assumptions \ref{as: U}--\ref{as: lambda} hold, and assume further that the reproducing kernels $\ka_X$, $\ka_M$ are bounded. If $\teps \succeq n^{-1/(2\teta+\frac{3\talpha+1}{2\talpha})}$, $\teps\succ n^{-1/2}$, and $n^{-1/2}\vartheta_n^Z[\tau_X(x)-\tau_X(x^*)]^{1/2} \succ \teps^{\teta}$, the asymptotic distribution for the NDE estimator is:
    $$
\sqrt{n}\left(\vartheta_n^Z[\tau_X(x)-\tau_X(x^*)]\right)^{-1/2}\left(\hNDE-\NDE\right)\overset{\c D}
    {\rightarrow} \c N(0, \E[W\otimes W]).
    $$
Let $\ttheta_n=n^{-1}\vartheta^Z_n[\Phi(x)-\Phi(x^*)]$, and  $ \theta_n=n^{-1}\vartheta^X_n[\tau_X(x)-\tau_X(x^*)]$. If  $\epsilon \succeq n^{-1/(2\eta+\frac{3\alpha+1}{2\alpha})}$, $\epsilon \succ n^{-1/2},\ttheta_n^{1/2} \succ \epsilon^{\eta}$, and $ \theta_n^{1/2} \succ \teps^{\teta}$, the asymptotic distribution for the NIE estimator is:
\begin{itemize}
    \item when $\frac{(2\teta+1)\talpha}{\talpha+1} <  \frac{(2\eta+1)\alpha}{\alpha+1}$, we have
    $\ttheta_n^{-1/2}\left(\hNIE-\NIE\right)\overset{\c D}
    {\rightarrow} \c N(0, \E[W\otimes W]);$
    \item when $\frac{(2\teta+1)\talpha}{\talpha+1} >  \frac{(2\eta+1)\alpha}{\alpha+1}$, we have
    $
    \theta_n^{-1/2}\left(\hNIE-\NIE\right)\overset{\c D}
    {\rightarrow} \c N(0, \E[(R_{VZ}U)\otimes (R_{VZ}U)]).
    $
\end{itemize}
\end{theorem}
The asymptotic normality of the direct effect estimator $\hNDE$ follows from Theorem \ref{thm: aympZ}. In contrast, the distributional complexity of $\widehat{\text{NIE}}$ arises from the sequential estimation of the two structural operators, $\hat{\Phi}$ and $\hat{\Psi}$, which may exhibit distinct convergence behaviors. Specifically, the operator characterized by slower spectral decay (measured by $(\alpha, \talpha)$) or a lesser degree of smoothness (measured by $(\eta, \teta)$) will incur a larger asymptotic variance, thereby dictating the overall convergence rate of the indirect effect.

\begin{remark}
   While \cite{sang_nonlinear_2022} establishes analogous asymptotic normality for functional data, our Theorem \ref{thm: aymp} offers two theoretical advancements. First, we achieve these convergence rates without invoking their restrictive Assumption 6 forcing approximation bias to be negligible relative to stochastic noise, which is frequently violated in non-parametric settings. Second, prior frameworks predominantly yield pointwise convergence, which is insufficient for global inference. In contrast, our weak convergence of $\hat{\Phi}$ enables a global test.
\end{remark}

\subsection{Pointwise Confidence Interval}\label{sec: CI}
Building on the asymptotic normality established in Theorem \ref{thm: aymcr}, we now construct computable pointwise confidence intervals for the causal effects. This requires consistent empirical estimators for both the scale functionals and the residual covariance operators. 

\begin{lemma}\label{lem: consistent est sigma}
    Suppose the conditions of Theorem \ref{thm: aymcr} hold. Let $f \in \mathcal{Z}$ and $g \in \mathcal{X}$ be elements with finite norms. Assuming that $\teps \succeq n^{-\frac{3}{4\teta+4}}$ and $\epsilon \succeq n^{-\frac{3}{4\eta+4}}$, we define $\hat\vartheta_n^Z[f] = \E_n[\langle \hSigzi f, \ka_Z(\dcd Z_i)-\hat\mu_Z\rangle^2]$ and $
    \hat\vartheta_n^X[g] = \E_n[\langle \hsigxi g,\ka_X(\dcd X_i)-\hat\mu_X\rangle^2].$
    If $\vartheta^Z_n[f] > 0$ and $\vartheta^X_n[g] > 0$, we have $\hat\vartheta^Z_n[f]/\vartheta^Z_n[f]\cP 1$ and $\hat\vartheta^X_n[g]/\vartheta^X_n[g]\cP 1.$
\end{lemma}
We must also consistently estimate the covariance operators of the structural residuals $W$ and $R_{VZ}U$. Let $\hat\Sigma_{R} = \frac{1}{n}\sum_{i=1}^n [\hat R_{VZ}(\tm(M_i) - \hat\Phi(X_i))]\otimes [\hat R_{VZ}(\tm(M_i) - \hat\Phi(X_i))]$. Inspired by \cite{liu2020}, who introduced a consistent estimator for error variance in linear ridge regression, we define the degree-of-freedom-adjusted empirical covariance operators: $
    \hat\Sigma_W = \frac{1}{n-\df}\sum_{i=1}^n Y_i \otimes \hat W_i$, where $\hat W_i = Y_i - \hat\Psi(M_i) - \hat\bgamma(X_i)$ and $\df = \tr(G_ZG_Z^\dagger + 1_n 1_n^T/n).$ 
\begin{lemma}\label{lem: Sig W R}
    Under Assumptions \ref{as: rho}--\ref{as: lambda}, suppose the kernels $\ka_X$ and $\ka_M$ are uniformly bounded. Then, $\hat{\Sigma}_W$ and $\hat{\Sigma}_{R}$ are consistent in the operator norm for $\E[W \otimes W]$ and $\E[(R_{VZ}U) \otimes (R_{VZ}U)]$, respectively. 
\end{lemma}
With these consistent estimates, we can construct asymptotically valid confidence intervals. For any $v \in \mathcal{H}$ and $q \in (0,1)$,  the $(1-q)$ confidence interval for $\langle \NDE, v\rangle$ is 
\begin{align}
    CI^{\NDE}(q, v) = \{y: |y - \langle\hNDE, v\rangle| \le \Phi_{\c N}^{-1}(1-q/2) \delta(v)\},\label{eq: CINDE}
\end{align}
where $\delta(v)^2=n^{-1}\hat\vartheta_n^Z[\tau_X(x)-\tau_X(x^*)]\langle \hat\Sigma_W v, v\rangle.$
By using the sum of the estimated variances from both the outcome and mediator models, for any $v \in \mathcal{H}$ and $q \in (0,1)$, the $(1-q)$ confidence interval for $\langle \NIE, v\rangle$ is given by: 
\begin{align}
    CI^{\NIE}(q, v) =& \{y: |y - 
    \langle\hNIE, v\rangle| \le \Phi_{\c N}^{-1}(1-q/2) \tilde\delta(v)\},\label{eq: CINIE}
\end{align}
where $\tilde \delta(v)^2=n^{-1}\hat\vartheta_n^Z[\hat\Phi(x)-\hat\Phi(x^*)]\langle \hat\Sigma_W v, v\rangle +n^{-1}\hat\vartheta_n^X[\tau_X(x)-\tau_X(x^*)]\langle \hat\Sigma_{R} v, v\rangle$. The construction of $\tilde \delta(v)^2$ bypasses the need to pre-test which asymptotic regime in Theorem \ref{thm: aymcr} dominates, as the asymptotically negligible variance term will naturally vanish.

\begin{proposition}\label{lem: CI}
        Under the conditions of Theorem \ref{thm: aymcr}, assume that $q \in (0,1)$, $\teps \succeq n^{-3/(4\tilde{\eta}+4)}$ and $\epsilon \succeq n^{-3/(4\eta+4)}$. As $n \to \infty$, both $CI^{\NDE}(q, v)$ and $CI^{\NIE}(q, v)$ achieve nominal coverage:
        $$P\left(\langle \NDE,v\rangle\in  CI^{\NDE}(q, v)\right)\rightarrow 1-q\quad \text{and} \quad P\left(\langle \NIE,v\rangle \in  CI^{\NIE}(q, v)\right)\rightarrow 1-q.$$
\end{proposition}

\subsection{Global Test Statistics} \label{sec: Global Tests}

\paragraph{Test for $H_0: \NDE=0$.} To test the global null hypothesis of no natural direct effect, we first consider the standardized estimation error for $\NDE$ in $\mathcal{H}$: 
$$
\hat{\c G}=\sqrt{n}\{\hat\vartheta_n^Z[\tau_X(x)-\tau_X(x^*)]\}^{-1/2}(\hNDE-\NDE).
$$
By Theorem \ref{thm: aymcr} and Lemma \ref{lem: consistent est sigma}, $\hat{\mathcal{G}}$ converges in distribution to a centered Gaussian random element $\mathcal{G}_W$ with covariance operator $\Sigma_W := \E[W \otimes W]$. Let $\{\lambda_j(A): j=1, \dots \}$  be the eigenvalues of an operator $A$, sorted in descending order. By Lemma S8 of the supplementary material, $\|\c G_W\|^2$ follows a weighted sum of independent chi-square distributions. Because $\|\c G_W\|^2$ is bounded in probability given the normality, it serves as a natural foundation for our global test. Under $H_0: \text{NDE} = 0$, the true parameter vanishes from the estimation error. Therefore, we define the test statistic as the squared norm of $\hat{\mathcal{G}}$ evaluated under $H_0$: 
$$T_{\NDE}=\langle \hat{\c G}, \hat{\c G}\rangle = n\{\hat\vartheta_n^Z[\tau_X(x)-\tau_X(x^*)]\}^{-1}\|\hNDE\|^2.$$

\begin{proposition}\label{prop: TNDE0}
    Given the conditions for Proposition \ref{lem: CI}, under $H_0: \text{NDE} = 0$, for any $t\in \R$, we have
    $
       P(T_{\NDE}\le t)-P(S \le t)\rightarrow 0, \text{ as } n \rightarrow \infty, 
    $
   where $S = \sum_{j=1}^\infty \lambda_j(\Sigma_W) \chi_{1,j}^2$, and $\{\chi^2_{1,j}\}_{j=1}^\infty$ are independent, standard chi-square random variables with one degree of freedom.
\end{proposition}

The asymptotic null distribution in Proposition \ref{prop: TNDE0} depends on the unknown eigenvalue spectrum $\{\lambda_j(\Sigma_W)\}_{j=1}^\infty$. To conduct inference in practice, we must approximate it using a computable surrogate for the residual covariance operator. Notably, the bias-corrected estimator $\hat{\Sigma}_W$ defined in Section \ref{sec: CI} is unsuitable for this specific task; because it was constructed via the cross-tensor products $Y_i \otimes \hat{W}_i$, it is not strictly symmetric. This asymmetry can lead to non-positive eigenvalues and numerical instability during spectral decomposition.  

To resolve this, we let $\hat{W}_i = Y_i - \hat{\Psi}(M_i) - \hat{\bgamma}(X_i)$ be the $i$-th empirical residual and define the symmetric surrogate as $\check{\Sigma}_W = \mathbb{E}_n[\hat{W} \otimes \hat{W}]$. By extracting the leading $l$ eigenvalues of $\check{\Sigma}_W$, we can form a truncated empirical estimate of the limiting random variable $S$.
\begin{theorem}\label{thm: TNDE}
   Given the conditions for Proposition \ref{lem: CI}, assume that $\E[\|W\|^4]$ is finite and $l^{1/2} \left(n^{-1/2} \teps^{-1}+ \teps^{\teta}\right) \rightarrow 0$ as $n, l \to \infty$. Under the null hypothesis $H_0: \text{NDE} = 0$, for any  $t\in \R$, we have 
     $
    P(T_{\NDE}\le t)-P(\check S^l_n\le t)\rightarrow 0, \text{ as } n,l \rightarrow \infty,
    $
    where $\check{S}^l_n=\sum_{j=1}^l \lambda_j(\check\Sigma_W) \chi_{1,j}^2$.
\end{theorem}
Theorem \ref{thm: TNDE} demonstrates that the quantiles of the target null distribution can be consistently approximated provided the truncation level $l$ is carefully controlled. In practice, we employ the Davies method \citep{daviesAlgorithmADavies} to approximate the distribution of $\sum_{j=1}^n \lambda_j(\check\Sigma_W) \chi_{1,j}^2$. By numerically inverting the characteristic function of the target distribution, the Davies method provides highly accurate and numerically stable approximations.

\paragraph{Test for $H_0: \NIE=0$.} To test the global null hypothesis of no natural indirect effect, we construct a pooled variance functional that spans both asymptotic regimes in Theorem \ref{thm: aymcr}. We define the empirical variance proxy as $\hat\varsigma^2_{\NIE} =  \hat\vartheta_n^Z[\hat \Phi(x) - \hat \Phi(x^*)]+\hat\vartheta_n^X[\tau_X(x)-\tau_X(x^*)].$ Depending on the specific asymptotic setting, this sum dynamically dominates and estimates either $\theta_n$ or $\tilde{\theta}_n$. Analogous to $T_{\text{NDE}}$, we define the squared-norm test statistic as: 
$$T_{\NIE} =n\hat\varsigma^{-2}_{\NIE}\| \hNIE\|^2.$$
Parallel to Theorem \ref{thm: TNDE}, we introduce an empirical operator for the random error covariance:
$$
\hat\Sigma^{\NIE}_n = \hat\varsigma^{-2}_{\NIE}\hat\vartheta_n^Z[\hat \Phi(x) - \hat \Phi(x^*)] \check\Sigma_W +\hat\varsigma^{-2}_{\NIE}\hat\vartheta_n^X[\tau_X(x)-\tau_X(x^*)]\hat\Sigma_R.
$$   

\begin{theorem}\label{thm: TNIE}
     Suppose the conditions of Proposition \ref{lem: CI} hold, and assume $\E[\|W\|_{\mathcal{H}}^4] < \infty$. Define $\tilde \varrho_n = n^{-1}\teps^{-2\teta}  \epsilon^{-(\alpha+1)/\alpha}$ and $\varrho_n =n^{-1}  \epsilon^{-2\eta}\teps^{-(\talpha+1)/\talpha}$. Under the null hypothesis  $H_0: \NIE=0$, assume that either of the following regime-specific conditions holds as $n, l \to \infty$:     
     
     (i) $\frac{(2\teta+1)\talpha}{\talpha+1} <  \frac{(2\eta+1)\alpha}{\alpha+1}$,
      $l^{1/2} \tilde\varrho_n \preceq 1$, and $l^{1/2} \left(n^{-1/2} \teps^{-1}+ \teps^{\teta}\right) \rightarrow 0$
     
     (ii) $\frac{(2\teta+1)\talpha}{\talpha+1} >  \frac{(2\eta+1)\alpha}{\alpha+1}$, $l^{1/2} \varrho_n \preceq 1$, and $l^{1/2} \left(n^{-1/2} \teps^{-1}+ \teps^{\teta}+ n^{-1/2} \epsilon^{-1}+ \epsilon^{\eta}\right) \rightarrow 0$. 
Then, 
    $
       \lim_{l,n\rightarrow \infty} P(T_{\NIE}\le t)-P(\tilde S_n^l\le t)\rightarrow 0 \text{ for any } t\in \R,
    $ where $\tilde S_n^l=\sum_{j=1}^l \lambda_j( \hat\Sigma^{\NIE}_n)\chi_{1,j}^2.$
\end{theorem}
As established in Theorem \ref{thm: aymcr}, the asymptotic behavior differs between cases (i) and (ii), determined by the unknown spectral decay and smoothness parameters $(\alpha, \eta, \talpha, \teta)$. However, as detailed in the proof of Theorem \ref{thm: TNIE}, either $\tilde{\varrho}_n$ or $\varrho_n$ converges to zero, ensuring that the bounds on $l$ can always be satisfied. Because $T_{\text{NIE}}$ and its finite-dimensional surrogate $\tilde{S}_n^l$ converge to the same distribution in both regimes, the procedure remains universally valid. This powerful property eliminates the need to distinguish between the two regimes or pre-estimate the unknown parameters $(\alpha, \eta, \tilde{\alpha}, \tilde{\eta})$ when implementing the test in practice.

\section{Simulation Study}\label{sec: simu}

We evaluate our proposed method across four simulated scenarios representing common non-Euclidean data types across four distinct causal pathway configurations in mediation analysis: (1) $\Phi, \bgamma,\Psi \neq 0$; (2) $\bgamma = 0$, and $\Phi, \Psi \neq 0$; (3) $\Psi = 0$, and $\Phi, \bgamma \neq 0$; (4) $\Phi = 0$, and $\Psi, \bgamma \neq 0$. Due to space constraints, we present Scenarios I and II below, deferring Scenarios III and IV to the supplement. Notably, Scenario I explores a novel setting featuring both a distributional mediator and a distributional outcome; to our knowledge, no existing literature accommodates this structure, precluding a baseline comparison. For Scenarios II--IV, we generate data utilizing both linear and nonlinear interactions and compare our method with DistM \citep{zhangetldistM}, FunM \citep{Coffman03092023}, and CCMM \citep{CompoMedianHongzhe}, respectively. The metric $d_M$ and $\langle\dcd  \cdot\, \rangle_{\c M}$ are tailored to each specific scenario. In the linear settings, $\ka_X$ and $\ka_M$ are linear kernels with $\ka_X(X_i, X_k) = X_i^T X_k$ and $\ka_M(M_i, M_k) = \langle M_i, M_k\rangle_{\c M}$. In the nonlinear settings, $\ka_M, \ka_X$ are Gaussian radial basis kernels $\ka_M(M_i, M_k) = \exp\left(-\gamma^M_G d^2_M(M_i, M_k)\right)$, and $\ka_X(X_i, X_k) = \exp\left(-\gamma^X_G (X_i - X_k)^2\right)$, where $(\gamma^X_G,\gamma^M_G)$ are selected by the GCV score over a fine grid as discussed in Section \ref{sec: est}.

\textbf{Scenario I: Distributional Mediator and Outcome.} We first evaluate our method when both the mediator and outcome take values in the Wasserstein-2 space (Remark \ref{rem: cod}). In practice, we observe $m$ independent realizations $\{M_{ij}\}_{j=1}^m$ for each true random distribution $M_i$. We approximate $M_i$ via the empirical measure $\hat{M}_i = \frac{1}{m}\sum_{j=1}^m \delta_{M_{ij}}$, allowing us to efficiently compute the empirical Wasserstein-2 distance using order statistics: $d_M(\hat M_i,\hat M_k) =  \{m^{-1}\sum_{j=1}^m (M_{i(j)}-M_{k(j)})^2\}^{1/2}$, where $M_{i(j)}$ is the $j$-th order statistic of the $i$-th sample. For $i=1,\dots, n$, we sample $X_i \sim \c N(0,1)$ and let $g(X_i) =  e/(1+\exp(-X_i^2))$, $h(X_i) =  -\exp(-X_i^2) + 2$. The mediator and outcome are generated as $M_i\sim\c N(\mu^M_i, 0.5^2)$ and  $Y_i\sim\c N(\mu^Y_i, \sigma^Y_i)$. We construct the mediator mean as $\mu^M_i = g(X_i) + U_i$, such that $U_i \sim \text{Uni}\left(-1, 1\right)$ for Settings I.1--I.3, and $\mu^M_i = U_i$ at Setting I.4. Letting $\mathcal{IG}(h_1, h_2)$ denote the inverse gamma distribution with mean $h_2/(h_1-1)$, the four causal pathway configurations are defined as:

\noindent \textbf{I.1 (full model)}: $\mu^Y_i = h(X_i) + \varsigma_i, \; \varsigma_i \sim \c N(0, 0.1^2), \quad \sigma^Y_i = h(X_i) + \langle M_i, \c N(0.7, 0.5)\rangle_{\c M}$

\noindent \textbf{I.2 (no NDE)}: $\mu^Y_i = -\langle M_i, \c N(1/4, 1)\rangle_{\c M} + \varsigma_i, \; \varsigma_i \sim \c N(0, 0.1^2), \quad \sigma^Y_i = \langle M_i, \c N(1/4, 1)\rangle_{\c M}$

\noindent \textbf{I.3 (no NIE)}: $\mu^Y_i = h(X_i) + \varsigma_i, \; \varsigma_i \sim \c N(0, 0.1^2), \quad \sigma^Y_i \sim h(X_i)\,\mathcal{IG}(16, 15)$

\noindent \textbf{I.4 (no NIE)}: $\mu^Y_i = \langle M_i, \c N(0.7, 0.5)\rangle_{\c M} + \varsigma_i, \; \varsigma_i \sim \c N(0, 0.1^2), \quad \sigma^Y_i = h(X_i)$

\begin{figure}[ht]
    \centering
    \includegraphics[width=0.9\linewidth]{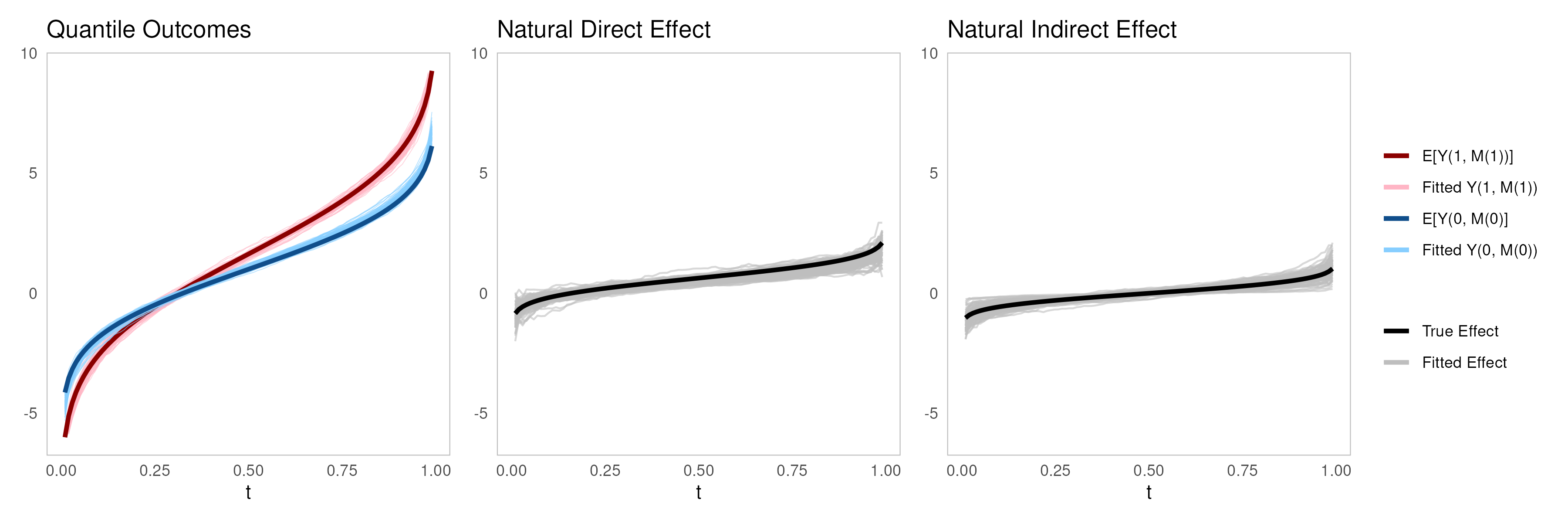}
    \caption{The estimated quantile functions for $\mathbb{E}[Y(0,M(0))]$ (blue lines) and $\mathbb{E}[Y(1,M(1))]$ (red lines), together with the natural direct effect (NDE) and natural indirect effect (NIE), across 100 simulation runs for the setting I.1. The true values are highlighted in bold.}
    \label{fig:distout est}
\end{figure}

\begin{figure}[htp!]
    \centering
    \includegraphics[width=1\linewidth]{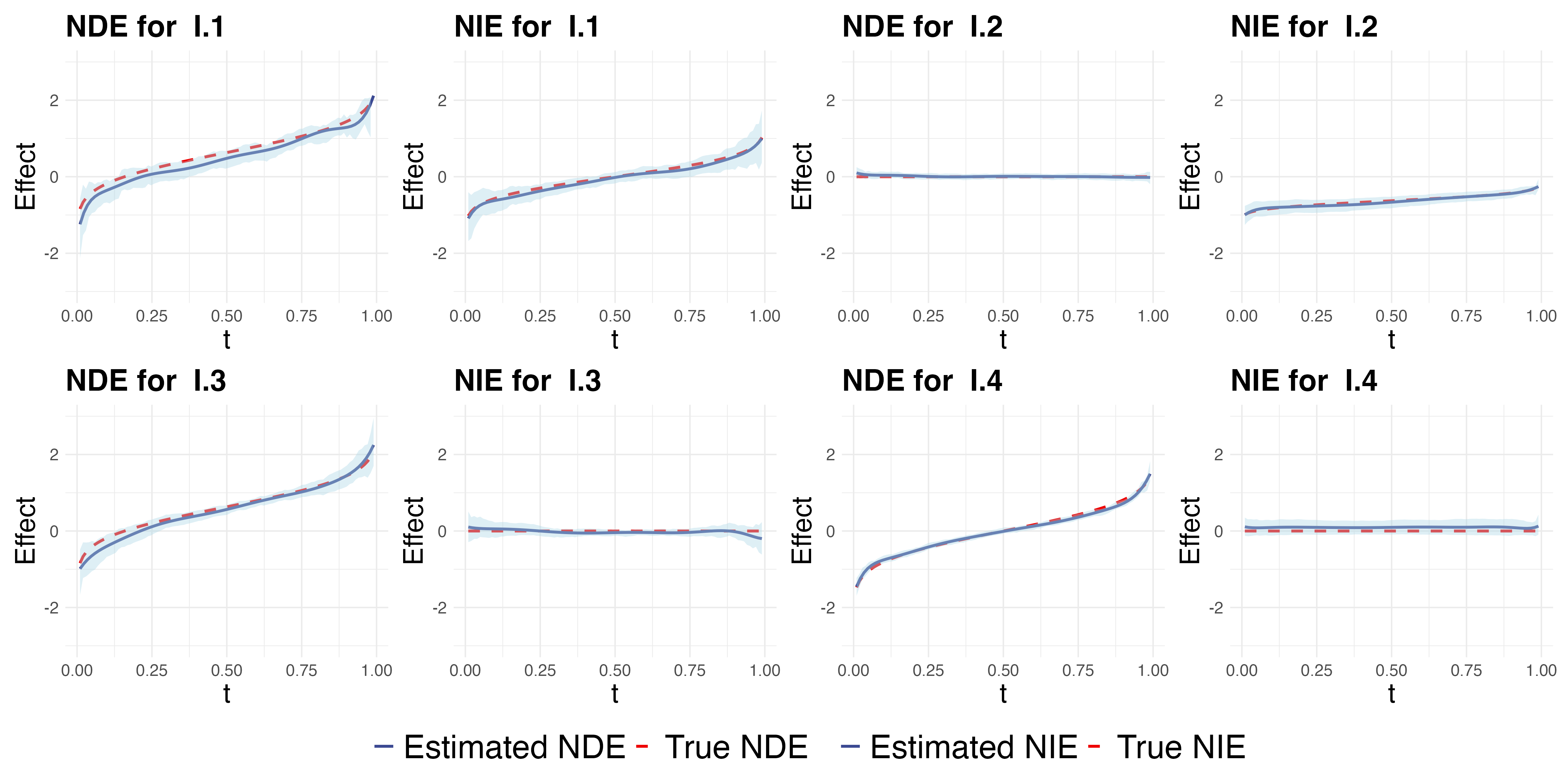}
    \caption{The estimated causal effects and pointwise confidence interval under setting I.1-I.4. The solid blue curve represents the estimated effect, the red dashed curve denotes the corresponding true effect, and the light blue band indicates the 95\% confidence interval. }
    \label{fig:distout pw_CI}
\end{figure}

\begin{figure}[ht]
    \centering
    \includegraphics[width=0.6\linewidth]{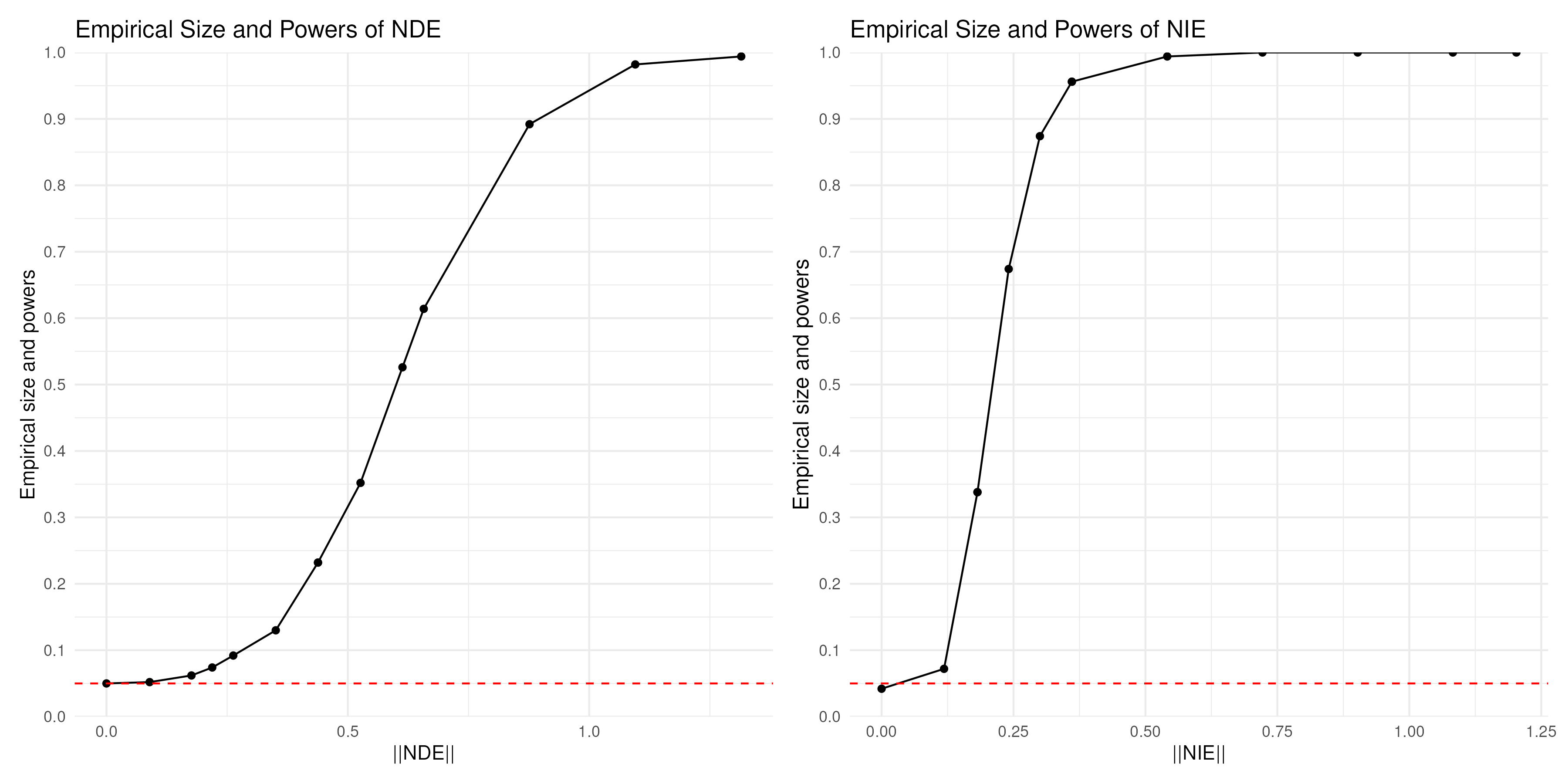}
    \caption{The empirical sizes and powers of our proposed model at size 0.05 at $l=n=100$ from 500 replications. The results are attained by varying the magnitude of the true value of $\|\NDE\|$ or $\|\NIE\|$. The red dotted line indicates the nominal type-I error rate of 0.05.}
    \label{fig:distout power}
\end{figure}

Figure \ref{fig:distout est} illustrates the estimation accuracy. The left panel demonstrates that the estimated quantile functions for the expected potential outcomes align with the ground truth. Similarly, the right panel confirms that the estimated NDE and NIE across 100 simulation runs accurately track the true causal effects. Consistent performance patterns for Settings I.2--I.4 are provided in Figure S1 of the supplement. 
Figure \ref{fig:distout pw_CI} displays the pointwise confidence bands for the NDE and NIE computed from a replication. Across all four settings, the estimated intervals consistently cover the true causal effects, empirically validating the coverage guarantees of Proposition \ref{lem: CI}.  Next, we evaluate the global testing procedure at a significance level of $0.05$. We use the data-generating process from Setting I.1, but vary the true magnitude $\|\NDE\|$ or $\|\NIE\|$ and compute the test statistics with $l=n=100$ over $500$ replicates. As shown in Figure \ref{fig:distout power}, the empirical sizes tightly control the Type-I error at the nominal $0.05$ level, while the power rapidly approaches $1$ as the true effect increases. We also examine the effect of parameter $l$. Additional sensitivity analyses in Figure S2 of the supplement confirm that the test remains robust when $l$ gets smaller.

\textbf{Scenario II: Distributional Mediator and Scalar Outcome.} This scenario examines the scalar exposure and outcome with a distributional mediator, retaining the Wasserstein-2 geometry ($\ka_M$, $d_M$, and $\langle \cdot, \cdot \rangle_{\mathcal{M}}$) in Scenario I. For $i=1,\dots, n$, we generate $X_i \sim \c N(0,1)$ and $M_i$ as a random distribution $\c N(\mu^M_i, \sigma^M_i)$, where $\sigma^M_i \sim 0.5\, \mathcal{IG}(4, 3)$. We define $
    g(X_i) = 2\sin(\pi X_i)+\exp(-X_i^2)$, and $  h(X_i) =0.5\sin(X_i)+\exp(-X_i^2)+ e\{1+\exp(-X_i^2)\}^{-1}$.
    
\noindent \textbf{II.1}: $\mu^M_i \sim \text{Laplace}(2X_i, 1); \; Y_i = -2X_i + \langle M_i, \c N(0.7, 0.5)\rangle_{\c M} + 1 + \tilde\varsigma_i, \; \tilde\varsigma_i \sim \c N(0, 0.1^2)$

\noindent \textbf{II.2}: $\mu^M_i \sim \text{Laplace}(X_i, 1); \; Y_i = \langle M_i, \c N(1, 1)\rangle_{\c M} + \tilde\varsigma_i, \; \tilde\varsigma_i \sim \c N(0, 0.5^2)$

\noindent \textbf{II.3}: $\mu^M_i \sim \text{Laplace}(X_i, 1); \; Y_i = -X_i + \tilde\varsigma_i, \; \tilde\varsigma_i \sim \c N(0, 0.5^2)$

\noindent \textbf{II.4}: $\mu^M_i \sim \c N(0, 0.2^2); \; Y_i = -2X_i + \langle M_i, \c N(0.7, 0.5)\rangle_{\c M} + 1 + \tilde\varsigma_i, \; \tilde\varsigma_i \sim \c N(0, 0.5^2)$

\noindent \textbf{II.5}: $\mu^M_i \sim \text{Laplace}(h(X_i), 1); \; Y_i = g(X_i) + \langle M_i, \c N(0.7, 0.5)\rangle_{\c M} + \tilde\varsigma_i, \; \tilde\varsigma_i \sim \c N(0, 0.1^2)$

\noindent \textbf{II.6}: $\mu^M_i \sim \text{Laplace}(e\{1+\exp(-X_i^2)\}^{-1}, 1); \; Y_i = \langle M_i, \c N(1, 1)\rangle_{\c M} + \tilde\varsigma_i, \; \tilde\varsigma_i \sim \c N(0, 1)$

\noindent \textbf{II.7}: $\mu^M_i \sim \text{Laplace}(h(X_i), 1); \; Y_i = g(X_i) + \tilde\varsigma_i, \; \tilde\varsigma_i \sim \c N(0, 0.5^2)$

\noindent \textbf{II.8}: $\mu^M_i \sim \c N(0, 0.2^2); \; Y_i = g(X_i) + \langle M_i, \c N(0.7, 0.5)\rangle_{\c M} + \tilde\varsigma_i, \; \tilde\varsigma_i \sim \c N(0, 0.5^2)$

\begin{figure}
    \centering
    \includegraphics[width=\linewidth]{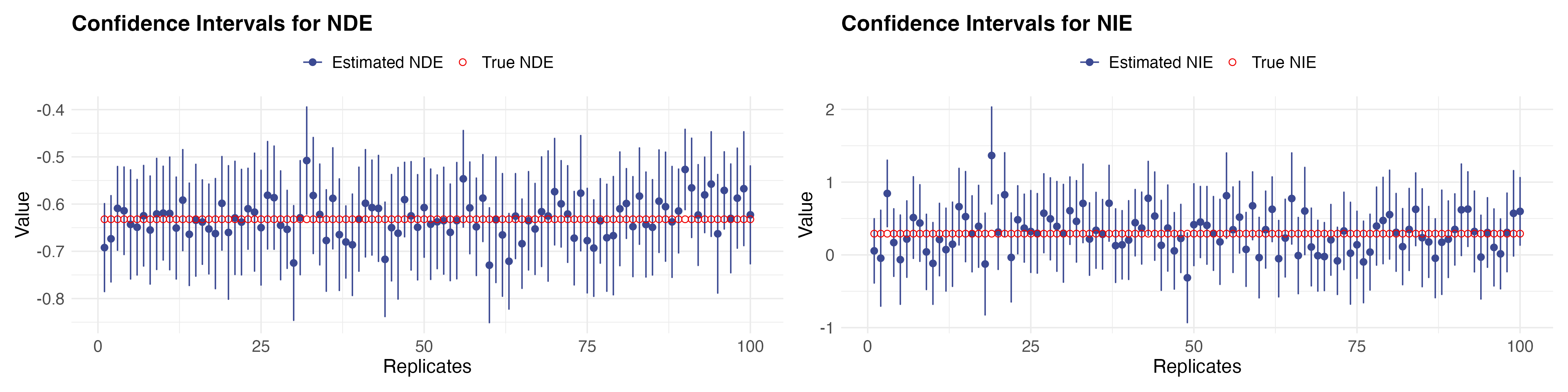}
    \caption{The 95\% confidence intervals under Scenario II.5 for 100 replicates.}
    \label{fig: CI_S2}
\end{figure}

Table \ref{tab: Sc2} reports the Mean Squared Errors (MSEs) across 100 replicates. In linear settings, ROMA achieves significantly lower prediction errors in II.1 and II.2, and performs comparably to DistM in II.3 and II.4 when mediators are absent. In nonlinear settings, ROMA substantially outperforms the strictly linear DistM. These results underscore ROMA's advantage in capturing complex mediation pathways. Figure \ref{fig: CI_S2} displays the estimated NDE and NIE confidence intervals for Setting II.5, demonstrating excellent finite-sample performance. The consistent coverage results for all other settings are provided in the supplement.

\begin{table}
\centering
{\small
\caption{Mean squared errors (MSE) of the Total Effect (TE), Natural Direct Effect (NDE), and Natural Indirect Effect (NIE) across 100  replicates, reported as mean (standard error).}
\label{tab: Sc2}
\renewcommand{\arraystretch}{0.65}
\small
\begin{tabular}{lllccc}
\toprule
\textbf{Type} & \textbf{Setting} & \textbf{Method} & \textbf{TE} & \textbf{NDE} & \textbf{NIE} \\
\midrule
\multirow{8}{*}{\textit{Linear}} 
  & \multirow{2}{*}{II.1} & ROMA & 0.012(0.002) & 0.000(0.000) & 0.012(0.002) \\
  &                       & distM    & 1.945(0.019) & 0.000(0.000) & 1.950(0.019) \\
  & \multirow{2}{*}{II.2} & ROMA & 0.032(0.004) & 0.004(0.001) & 0.029(0.004) \\
  &                       & distM    & 1.000(0.020) & 0.003(0.001) & 0.988(0.016) \\
  & \multirow{2}{*}{II.3} & ROMA & 0.006(0.001) & 0.004(0.001) & 0.005(0.001) \\
  &                       & distM    & 0.007(0.002) & 0.004(0.000) & 0.003(0.001) \\
  & \multirow{2}{*}{II.4} & ROMA & 0.004(0.000) & 0.003(0.000) & 0.001(0.000) \\
  &                       & distM    & 0.004(0.001) & 0.003(0.000) & 0.002(0.001) \\
\midrule
\multirow{8}{*}{\textit{Nonlinear}} 
  & \multirow{2}{*}{II.5} & ROMA & 0.080(0.014) & 0.002(0.000) & 0.075(0.013) \\
  &                       & distM    & 0.190(0.014) & 0.474(0.020) & 0.082(0.004) \\
  & \multirow{2}{*}{II.6} & ROMA & 0.195(0.027) & 0.047(0.015) & 0.154(0.020) \\
  &                       & distM    & 0.415(0.017) & 0.010(0.001) & 0.385(0.010) \\
  & \multirow{2}{*}{II.7} & ROMA & 0.001(0.000) & 0.001(0.000) & 0.000(0.000) \\
  &                       & distM    & 0.500(0.022) & 0.474(0.020) & 0.004(0.001) \\
  & \multirow{2}{*}{II.8} & ROMA & 0.006(0.001) & 0.002(0.000) & 0.003(0.001) \\
  &                       & distM    & 0.471(0.022) & 0.472(0.019) & 0.005(0.002) \\
\bottomrule
\end{tabular}
}
\end{table}

\section{Applications}\label{sec: RDA}

\noindent\textbf{Application to Microbiome Data.} Diet influences human health partly by altering the gut microbiome \citep{COMBO2011Gary}. Using data from the COMBO study, we investigate whether the microbiota mediates the effect of long-term fat intake on body mass index (BMI). From $3,068$ initial OTUs, we aggregated to the genus level, retained the 45 genera present in at least 10\% of samples, and obtained $n=96$ filtered subjects. 
We apply ROMA and CCMM \citep{CompoMedianHongzhe}, defining standardized fat intake as the exposure ($X$), the 45-dimensional compositional vector as the mediator ($M$), and BMI as the outcome ($Y$). We adopt the same model construction for ROMA as in the nonlinear settings of Scenario IV. The estimates and confidence intervals are reported in Table \ref{tab: RDA micro}. At $\alpha = 0.05$, neither method detects significant effects. However, at $\alpha = 0.10$, our method identifies both the NDE and NIE as significant, whereas CCMM only detects a significant NDE. Biologically, the gut microbiota influences adiposity by modulating dietary energy extraction \citep{Turnbaugh2006}, making an indirect effect highly plausible. While CCMM likely lacks the statistical power to detect this NIE, given the moderate sample size ($n=96$), our framework successfully recovers it. Furthermore, Table S6 in the supplement shows our approach substantially outperforms CCMM, yielding biologically consistent and empirically robust estimates.

\begin{table}[htp]
\centering
{\small
\caption{Mediation analysis results for ROMA and CCMM with confidence intervals.}
\label{tab: RDA micro}
\renewcommand{\arraystretch}{0.7}
\begin{tabular}{llccc}
\toprule
\textbf{Effect} & \textbf{Method} & \textbf{Estimate} & \textbf{90\% CI} & \textbf{95\% CI} \\
\midrule
\multirow{2}{*}{NDE} & ROMA & 1.146 & [0.161, 2.131] & [$-0.028$, 2.320] \\
                     & CCMM     & 0.909 & [0.173, 1.662] & [$-0.054$, 1.738] \\
\multirow{2}{*}{NIE} & ROMA & 2.014 & [0.098, 3.931] & [$-0.269$, 4.298] \\
                     & CCMM     & 0.813 & [$-0.194$, 2.056] & [$-0.389$, 2.252] \\
\bottomrule
\end{tabular}
}
\end{table}

\begin{figure}
    \centering
    \includegraphics[width=\linewidth]{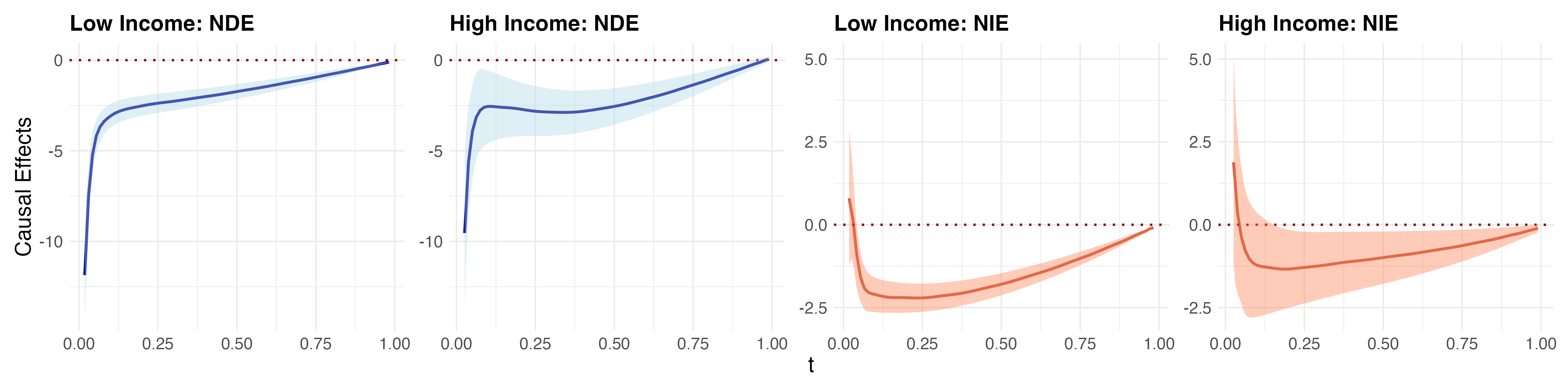}
     \caption{The estimated causal effects and 95$\%$ pointwise confidence intervals. 
     }
    \label{fig: air pollution est}
\end{figure}

\noindent\textbf{Application to Air Pollution and Mortality.}
Extreme temperatures and air pollution are significant contributors to global mortality \citep{zanobetti2008temperature, wu2020air}. Because the impact of heat waves on mortality is intensified on high-pollution days \citep{analitis2014effects}, we investigate whether air pollution mediates the relationship between heat and human health using data from $n=2231$ U.S. counties on the CDC website. We constructed cohort life tables transformed into age-at-death quantiles as the distributional outcome, with annual average PM2.5 estimated as in \cite{josey2023air} as the mediator and summer temperature as the exposure. Following \cite{josey2023air}, we stratified the counties by median household income into $558$ high-income and $1,673$ low-income counties. Quantile outcomes were embedded in $L^2[0,1]$ with linear kernels. Reference exposures were chosen as the 5th and 95th percentiles. As shown in Figure~\ref{fig: air pollution est}, the NDE is significantly negative across $[0,1]$ for both groups, confirming the direct harmful impact of heat on longevity. However, a disparity emerges in the mediation effects: the global test identifies both NDE and NIE as highly significant in low-income counties ($p=0.003$), whereas the NIE is negligible in high-income counties ($p=0.369$). These results suggest that high temperatures adversely affect health directly and indirectly through PM2.5 in economically disadvantaged populations, highlighting significant climate-related health inequities linked to socioeconomic status.

\section{Conclusion}\label{sec: conclusion}
  {We propose a novel and unified framework for nonlinear causal mediation analysis where the exposures, mediators, and outcomes are random objects in metric spaces.} By embedding these variables into RKHS and general Hilbert spaces, we capture complex linear and nonlinear relationships via linear Hilbertian regressors. We derive the asymptotic distributions of our causal estimators, enabling both pointwise and global inference. Real-world applications demonstrate the framework's capacity to uncover significant biological pathways and socioeconomic health inequities that traditional methods miss.

{\small
\renewcommand{\baselinestretch}{1}
\bibliographystyle{agsm}
\bibliography{ref}
}
\end{document}

%% file: macros.tex
\def\boe{\begin{enumerate}}
\def\eoe{\end{enumerate}}

\newtheorem{proposition}{{\bf Proposition}}
\newtheorem{lemma}{{\bf Lemma}}
\newtheorem{corollary}{{\bf Corollary}}
\newtheorem{assumption}{{\bf Assumption}}
\newtheorem{definition}{{\bf Definition}}
\newtheorem{theorem}{{\bf Theorem}}

\newtheorem{example}{{\bf Example}}

\newtheorem{remark}{{\bf Remark}}
\newcommand\ca[1]{{\cal{#1}}}
\newcommand\lo[1]{_{\nano{#1}}}

\def\Om{\Omega_M}
\def\Oy{\Omega_Y}
\def\bbeta{\boldsymbol \beta}
\def\bgamma{\boldsymbol \gamma}

\def\tm{\tau_M}
\def\c{\mathcal}
\def\Eo{\E_{\oplus}}

\def\dcd{\,\cdot\,,\,}
\def\Sigz{\Sigma_{ZZ}}
\def\Sigx{\Sigma_{XX}}
\def\Sigm{\Sigma_{MM}}

\def\hSigzi{\hat\Sigma_{ZZ}^\dagger}
\def\hsigxi{\hat\Sigma_{XX}^\dagger}
\def\Sigzv{\Sigma_{ZV}}
\def\hSigz{\hat{\Sigma}_{ZZ}}
\def\hSigx{\hat{\Sigma}_{XX}}

\def\ttheta{\tilde{\theta}}
\def\teta{\tilde{\eta}}
\def\teps{\tilde{\epsilon}}
\def\talpha{\tilde{\alpha}}

\def\cP{\overset{P}{\rightarrow}}
\def\NDE{\operatorname{NDE}}
\def\hNDE{\widehat{\operatorname{NDE}}}
\def\NIE{\operatorname{NIE}}
\def\hNIE{\widehat{\operatorname{NIE}}}

\def\argmin{\mathrm{argmin}}

\def\E{\mathbb{E}}
\def\R{\mathbb R}

\def\L{{\cal L}}

\def\var{\mathrm{var}}
\def\cov{\mathrm{cov}}

\newcommand{\indep}{\;\, \rule[0em]{.03em}{.65em} \hspace{-.41em}
\rule[-.02em]{.65em}{.03em} \hspace{-.41em}
\rule[0em]{.03em}{.65em}\;\,}

\def\tr{\mathrm{tr}}

\def\ran{\mathrm{ran}}

\def\nano{\scriptscriptstyle}

\def\ka{\kappa}

\def\L2T{L \lo 2 (T)}
\def\L2TX{L \lo 2 (T\lo X)}
\def\L2TX{L \lo 2 (T\lo Y)}

\def\eod{\end{document}}

\def\shortbar{\rule[-.0em]{.04em}{.15em}}
\def\bbar{\mbox{\raisebox{-.05em}{$\,\overset{\overset{\shortbar}{\shortbar}}{\shortbar}\,$}}}

\newfont{\rsfsten}{rsfs10 scaled 1100}
\newfont{\rsfstena}{rsfs10 scaled 800}
\newfont{\rsfstenb}{rsfs10 scaled 500}

\def\shortbar{\rule[-.0em]{.04em}{.15em}}
\def\bbar{\mbox{\raisebox{-.05em}{$\,\overset{\overset{\shortbar}{\shortbar}}{\shortbar}\,$}}}